%% file: ms.tex
\documentclass[nojss]{jss}

\usepackage[utf8]{inputenc}

\usepackage{thumbpdf,lmodern}

\usepackage{framed}

\usepackage{hyperref}
\usepackage{soul}
\usepackage[labelfont=bf]{caption}
\usepackage{subcaption}
\usepackage{tikz}
\usepackage{amsmath}
\usetikzlibrary{arrows.meta}
\tikzset{%
  >={Latex[width=2mm,length=2mm]},
            base/.style = {rectangle, rounded corners, draw=black,
                           minimum width=5cm, minimum height=1cm,
                           text centered, font=\sffamily},
  standard_style/.style = {base, fill=white}}

\author{Christian Winkler\\University of Bonn
   \And Katharina Linden\\University of Bonn
   \And Andreas Mayr\\University of Bonn
   \And Thomas Schultz\\University of Bonn
   \AND Thomas Welchowski\\University of Bonn
   \And Johannes Breuer\\University of Bonn
   \And Ulrike Herberg\\University of Bonn}
\Plainauthor{Christian Winkler}

\title{RefCurv: A Software for the Construction of Pediatric Reference Curves}
\Plaintitle{RefCurv: A Software for the Construction of Pediatric Reference Curves}
\Shorttitle{RefCurv: A Software for the Construction of Pediatric Reference Curves}

\Abstract{
In medicine, reference curves serve as an important tool for everyday clinical practice. Pediatricians assess the growth process of children with the help of percentile curves serving as norm references. The mathematical methods for the construction of these reference curves are sophisticated and often require technical knowledge beyond the scope of physicians. An easy-to-use software for life scientists and physicians is missing. As a consequence, most medical publications do not document the construction properly.
This project aims to develop a software that enables non-technical users to apply modern statistical methods to create and analyze reference curves.\\
In this paper, we present RefCurv, a software that facilitates the construction of reference curves.
The software comprises functionalities to select and visualize data. Users can fit models to the data and graphically present them as percentile curves. Furthermore, the software provides features to highlight possible outliers, perform model selection, and analyze the sensitivity. \\
RefCurv is an open-source software with a graphical user interface (GUI) written in \proglang{Python}. It uses \proglang{R} and the \pkg{gamlss} add-on package (\cite{gamlss2005}) as the underlying statistical engine.\\
RefCurv simplifies the process to create percentile curves with a broad set of data processing features. The tool can help to standardize the procedure and plan the acquisition of data. 
An exemplary analysis of the robustness for the underlying statistical methods is shown in case scenarios. Also, a method to design studies concerning the required sample size and the model setting is demonstrated.\\
In summary, RefCurv is the first software based on the \pkg{gamlss} package, which enables practitioners to construct and analyze reference curves in a user-friendly GUI. In broader terms, the software brings together the fields of statistical learning and medical application. Consequently, RefCurv can help to establish the construction of reference curves in other medical fields.

}

\Keywords{reference curves, percentile curves, centile estimation, z-scores, LMS method, pediatrics, echocardiography, open source, \proglang{R}, \pkg{gamlss}, \proglang{Python}}
\Plainkeywords{}

\Address{
  Christian Winkler\\
  Department of Pediatric Cardiology\\
  Center for Pediatrics\\
  University of Bonn\\
  Adenauerallee 119\\
  D-53113 Bonn\\
  E-mail: \email{christian.winkler@ukbonn.de}\\
}

\begin{document}


\newpage
\section[Introduction]{Introduction} \label{sec:intro}
Reference curves and charts are standard tools to describe the normal range of a parameter. In clinical practice, physicians use percentile curves (or z-score curves) to evaluate measured values of patients. Comparing the measurement to a reference helps to quantify the severity of the disease and diagnose the condition of a patient. In this context, percentile curves have been established for most common physiological and anthropometric parameters. For children, the curves can be used to assess the growth process. In literature, several reference curves and charts for pediatric parameters are available. One prominent parameter is the Body Mass Index (BMI) (\cite{Cole1995, fredriks2000}).\\

\begin{figure}[htbp]
\centering
\includegraphics[width=0.65\textwidth]{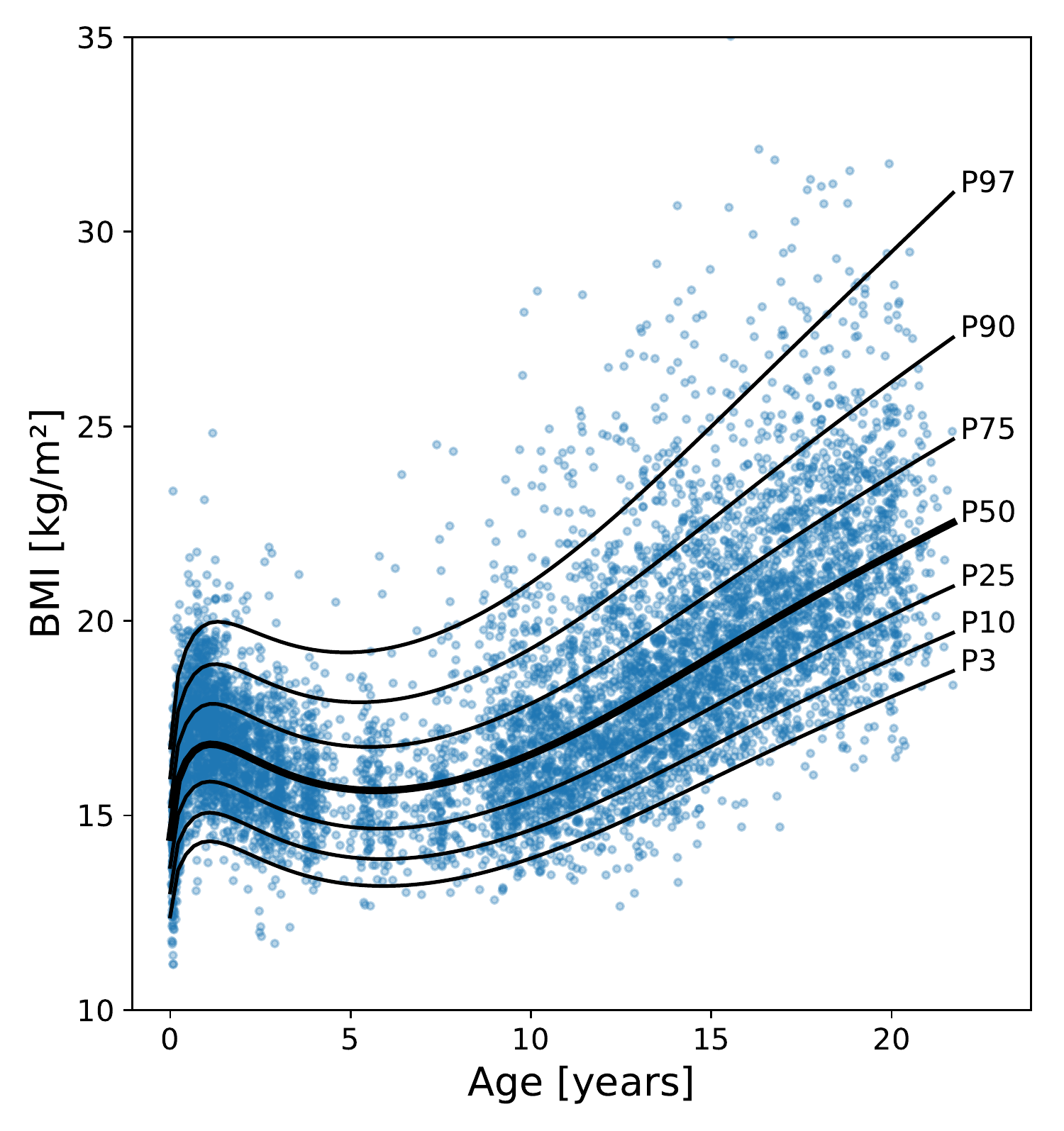}
\caption{\label{fig:bmiExample} \textbf{Reference curves for BMI based on a dataset of healthy Dutch boys (\cite{fredriks2000continuing}).} RefCurv was used to fit a model to the data points and depict it in the form of percentile curves. The labels indicate the percentiles, e.g. "P3" stands for the third percentile.}
\end{figure}

Figure \ref{fig:bmiExample} shows an example of pediatric reference curves for the BMI, which we fitted to a dataset from a previous study (\cite{fredriks2000continuing}). Furthermore, there have been studies on weight, height and head circumference (\cite{Cole1998, WHOMulticentregrowthreferencestudygroup2006, Cacciari2006, neuhauser2013referenzperzentile}).
\\

Another broad application field for reference curves are echocardiographical parameters \\ (\cite{kobayashi2016, dallaire2011, cantinotti2017}). Echocardiography has become an essential support for cardiological examination in children. Due to its noninvasiveness and fast application, it has been established as a standard technology in everyday clinical practice. Cardiologists use reference curves to detect cardiac pathologies and plan surgical treatments. A recent literature review on this growing field is given by \cite{mawad2013}. In the present paper, the focus is on the cardiological application of our software in children and examples are based on echocardiographical measurements.\\

The mathematical methods for the construction of pediatric reference curves have been shaped by the publications of Cole and Green (\cite{Cole1990, Cole1992, Cole1998}). In \cite{Cole1990}, the author proposes an algorithm for fitting smooth curves to data by using a penalized likelihood. Furthermore, Cole and Green describe the Box-Cox Cole Green (BCCG) distribution for pediatric growth curves and show the application on a dataset for BMI (\cite{Cole1995}). This approach has since been called the LMS method (or Lambda-Mu-Sigma method) and has been applied in many studies (\cite{mul2001pubertal, fredriks2000continuing, katzmarzyk2004waist, nysom2001body, ataei2016blood, hirschler2016waist, khadilkar2014waist}).\\

Subsequently, a program called LMSchartmaker was implemented by the group of Cole and Green enabling practitioners to apply the LMS method. This tool was used by multiple studies but we found issues regarding the scientific practice. On the one hand, LMSchartmaker is not open-source and there is not any description of the implementation. On the other hand, 
a scientific publication and references are missing.\\
At the same time, Rigby and Stasinopoulos developed and implemented the \proglang{R} add-on package for "Generalized Additive Models for Location Scale and Shape" (\pkg{gamlss}, \cite{gamlss2005, gamlss2007}). The \pkg{gamlss} package contains the LMS method and algorithms by Cole and Green. In addition, it extends the method by providing other model classes and diagnostic tools to assess the fitted reference curves. Unlike LMSchartmaker, the \pkg{gamlss} package is open-source, scientifically well-documented and free of charge. However, the usage of \pkg{gamlss} requires an intense study of related mathematical methods and programming skills in \proglang{R}.\\
Despite the availability of multiple statistical methods, most medical publications do not document the construction of reference curves properly or miss important information such as details about the model selection. One reason might be the complex application of statistical methods, which is a challenge for physicians and data analysts alike.\\ Researchers often cannot reproduce study results in the form of reference curves because datasets are not published by the authors.\\

\newpage
This project aims to develop RefCurv, an easy-to-use software for the construction of reference curves. With this tool, we want to enable non-technical users to create and analyze percentile curves for clinical usage. Moreover, it was intended to help experts with the advanced analysis of reference curves. Likewise, we proposed features to plan the study design such as estimating the required sample size. \\

RefCurv uses \proglang{R} and the \pkg{gamlss} add-on package as the underlying statistical engine. Users can apply the LMS method on their data or use a customized GAMLSS model for their calculations. The graphical user interface (GUI) is written in Python using its features in data visualization and processing. Users can define model settings in the GUI. This information is passed on to an \proglang{R}-script using functions from the \pkg{gamlss} package. After computation, the results are delivered back to the GUI. Functionalities for data selection, model selection, and model validation are provided. The software is designed to simplify data processing and model fitting. With RefCurv, users are guided through a simple workflow from acquired data to reference curves. The software is intended mainly for users without any specific programming or mathematical skills. \\
Furthermore, an echocardiographical dataset was acquired in previous studies by our research group. We use these data to demonstrate and explain the application of RefCurv. The given examples can be considered as recommended steps for the construction of reference curves.\\
 

\newpage
\section[Methods]{Methods} \label{sec:methods}
The main focus of RefCurv lies on the LMS method by Cole using the \pkg{gamlss} package in \proglang{R} for the statistical computations. The model used for the LMS method is a special case of a GAMLSS model. The model class is defined by the Box-Cox Cole Green (BCCG) distribution and penalized splines as smoothing for the distribution parameters L, M and S. Penalized splines are implemented in \proglang{R} as \code{pb()} function. Each penalized spline has a degree of freedom (\code{df}) to be predefined by the user. These three parameters (\code{L_df}, \code{M_df} and \code{S_df}) are arguments of the \code{pb()} function and they are therefore defined as hyperparameters. We chose a setting of \code{L_df} = 0, \code{M_df} = 1 and \code{S_df} = 0 as the default model.\\
Consequently, the model fitting according to the LMS method is implemented as followed:
\begin{CodeChunk}
\begin{CodeInput}
LMS_model <- gamlss(y ~ pb(x, df = M_df),
		    sigma.formula = ~ pb(x, df = S_df),
		    nu.formula = ~ pb(x, df = L_df),
		    family = "BCCG",
		    method = RS(),
		    data = dataset_training)
\end{CodeInput}
\end{CodeChunk}
The Rigby and Stasinopoulos algorithm, \code{RS()}, is used for the fitting \citep{gamlss2007}.\\

The LMS method has been established as a standard procedure for pediatric reference curves. Therefore, it is set as default for the model fitting (Appendix \ref{app:lms}). Apart from that, the advance settings in RefCurv allow the user to fit a broader set of univariate GAMLSS models to the data.\\
Details about the installation and software architecture of RefCurv are given in the appendix (Appendix \ref{app:refcurv}). The statistical engine is the \pkg{gamlss} package and RefCurv consequently inherits its limitations.\\
RefCurv's model fitting is based on a model class with a BCCG distribution. One limitation is that this model class is developed and tested for positive data values only. Furthermore, methods might be sensitive to outliers and model fitting might fail if data is distributed unevenly. These limitations and how to address them will be discussed in the application section. \\
In this section, we will describe RefCurv in four parts. After giving information about the repository and documentation (\ref{ch21}), we will present its graphical user interface (\ref{ch22}). Next, RefCurv's features and functions are presented (\ref{ch23}). Finally, we will recommend steps for the construction of reference curves (\ref{ch24}).

\subsection{GitHub repository and documentation} \label{ch21}
RefCurv is open source and currently available as version 0.4.2. The source code and binaries are provided on GitHub: \href{https://github.com/xi2pi/RefCurv}{https://github.com/xi2pi/RefCurv}\\
The related \href{https://github.com/xi2pi/RefCurv/wiki}{GitHub Wiki} contains a quick guide and instructions for the application. The example datasets, which were used in this paper, can be accessed through the software directly. The repository will be kept up to date and news will be announced on GitHub. Developers can exchange information in the related \href{https://github.com/xi2pi/RefCurv/issues}{forum for issues}.
\newpage
In addition to the source code, we created a website (\href{https://refcurv.com}{https://refcurv.com}) and video tutorials \\(\href{https://vimeo.com/user93523411}{https://vimeo.com/user93523411}).\\
RefCurv has been developed and tested on Windows and Linux. More information about package versions are given in Appendix \ref{app:refcurv}.
\subsection{Graphical user interface} \label{ch22}
\begin{figure}[htbp]
\centering
\includegraphics[width=1.0\textwidth]{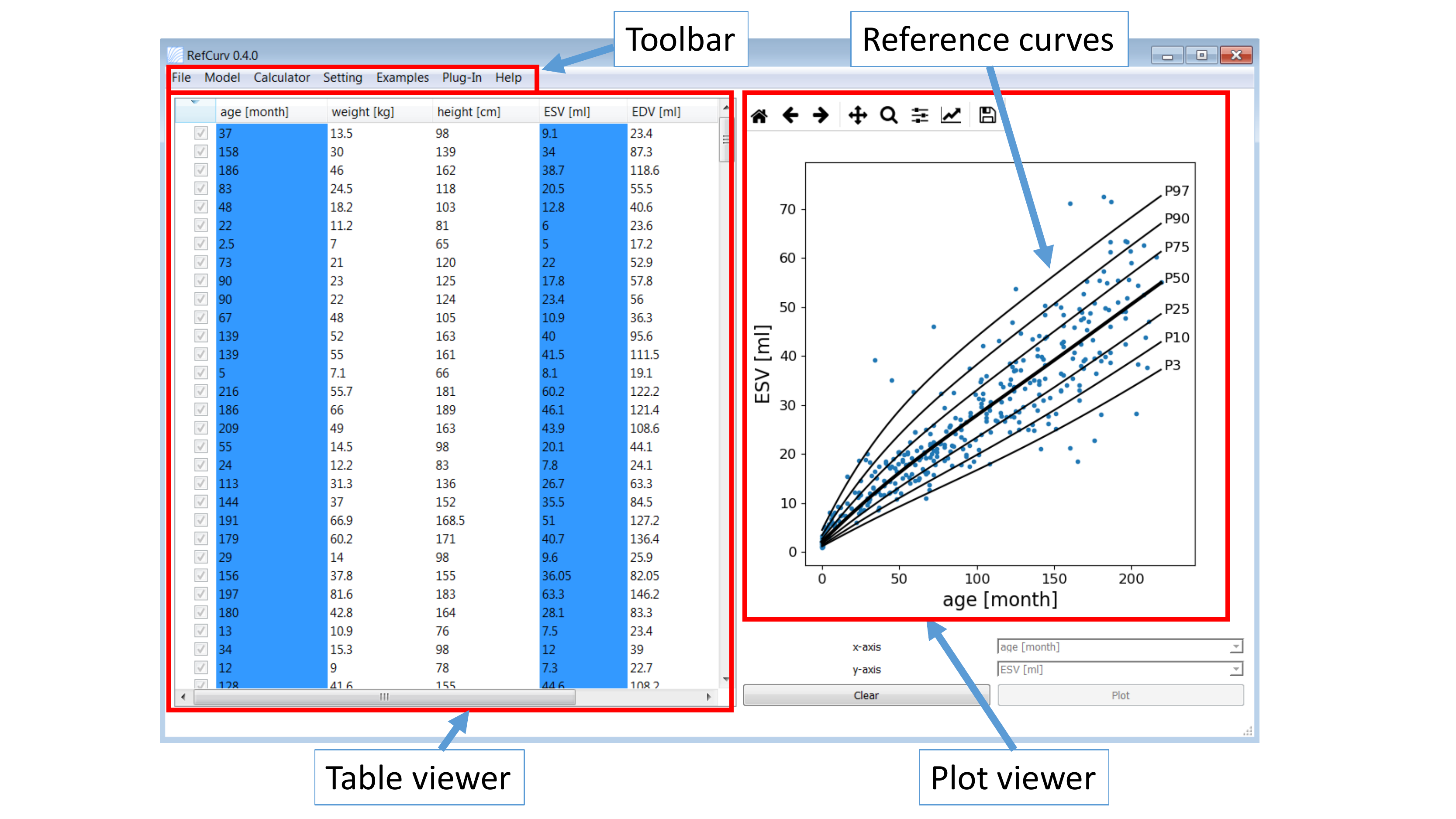}
\caption{\label{fig:refcurv_GUI} \textbf{RefCurv's graphical user interface with a table viewer (left) and a plot viewer (right).} Users can navigate through the features by using the upper toolbar.}
\end{figure}

Figure \ref{fig:refcurv_GUI} shows RefCurv's graphical user interface (GUI) consisting of a table viewer (left) and a plot viewer (right). Users can select data columns in the table viewer and visualize them in the plot viewer as scatter plot. In this example, we demonstrate the application on an echocardiographical dataset. The end-systolic volume of the left ventricle (ESV) is plotted against the age. The default model was fitted and graphically depicted as curves in the plot viewer. Each curve represents the percentile of the underlying distribution (3rd, 10th, 25th, 50th, 75th, 90th, 97th) and is labeled accordingly (e.g. "P3" stands for the third percentile). Percentile curves can be easily converted into z-score curves.\\
Users can navigate in RefCurv through the toolbar at the top of the window. The toolbar consists of a set of buttons for categories such as "Model" for the model processing. RefCurv is a Multiple Document Interface application, meaning that users can adjust settings in subwindows for most functions. 

\newpage
\subsection{Features} \label{ch23}
\subsubsection{Import of data}
RefCurv allows the import of data tables in the form of CSV files ("File" $\rightarrow$ "Load Data"). The following structure of the data table is required: columns contain measured variables, while rows represent the cases. The first row of the chart should be a header indicating the name of the measured variables.
\begin{table}[h]
\centering
\begin{tabular}{|l||c|c|}
\hline 
Subjects & Variable 1 & Variable 2 \\ 
\hline 
\hline
Subject 1 & ... & ... \\ 
\hline 
Subject 2 & ... & ... \\ 
\hline 
... & ... & ... \\ 
\hline 
\end{tabular}
\caption{\textbf{Structure of the input table for RefCurv}} 
\end{table}

After import, users can inspect the data in the table viewer as highlighted in Figure \ref{fig:refcurv_GUI}.
\subsubsection{Data selection}
After the data are loaded, RefCurv will provide functions to select data points and exclude them in case they are considered as anomalies. For that, users can choose two variables in the lower right drop down menu, one for the x-axis and one for the y-axis. Chosen columns are highlighted in the table viewer. By clicking the "Plot" button, a scatter plot is created. Data points are highlighted in the scatter plot when chosen in the table viewer. By checking or unchecking the box in the table, subjects can be excluded or included respectively. Chosen data serves as the training dataset and can be used for the model fitting.
\begin{figure}[htbp]
\centering
\includegraphics[width=0.8\textwidth]{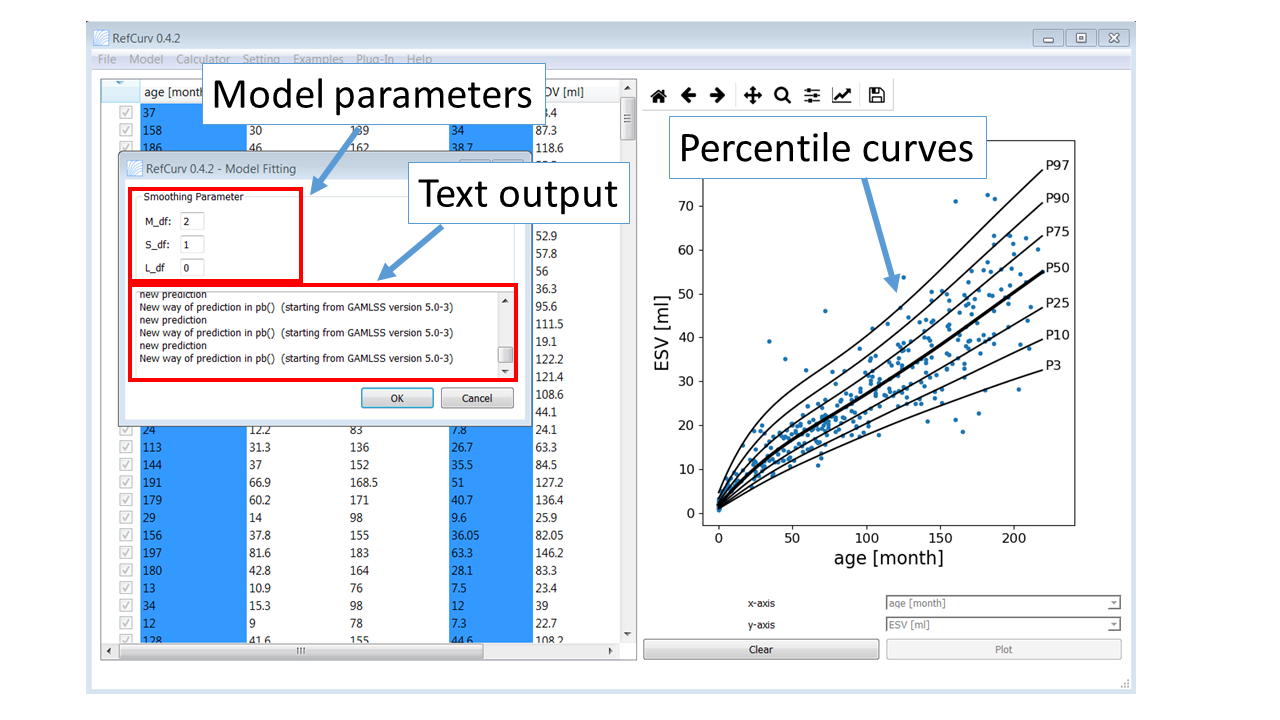}
\caption{\label{fig:refcurv_model_fit}\textbf{RefCurv's model fitting.} The model fitting window shows the model parameters and gives a summary in the text output field.}
\end{figure}

\subsubsection{Model fitting}
\begin{figure}[htp]

\centering
\begin{subfigure}[c]{.3\textwidth}
\includegraphics[width=1\textwidth]{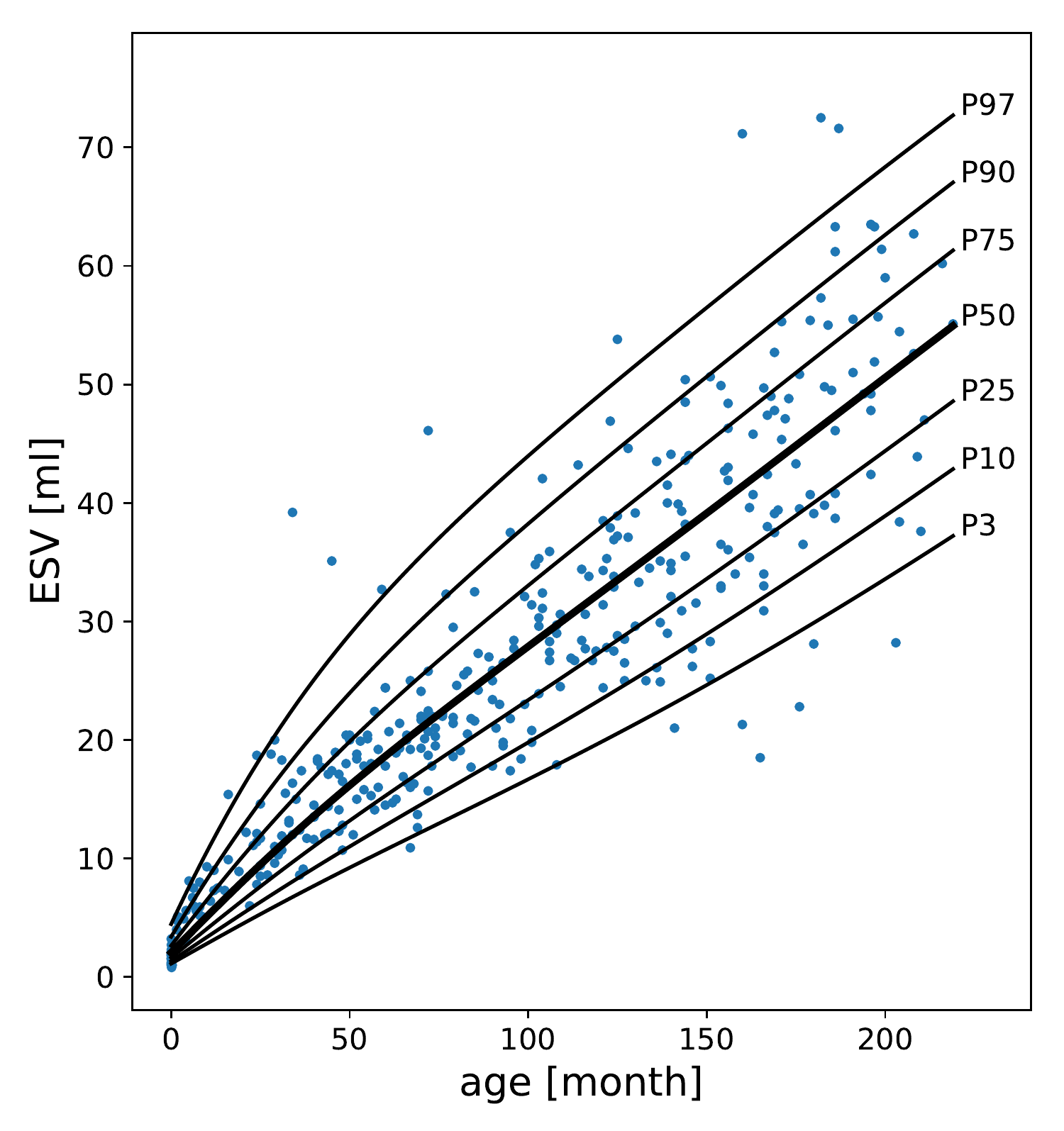}
\subcaption{\scriptsize\code{L\_df} = 0, \code{M\_df} = 1, \code{S\_df} = 0}
\end{subfigure}
\begin{subfigure}[c]{.3\textwidth}
\includegraphics[width=1\textwidth]{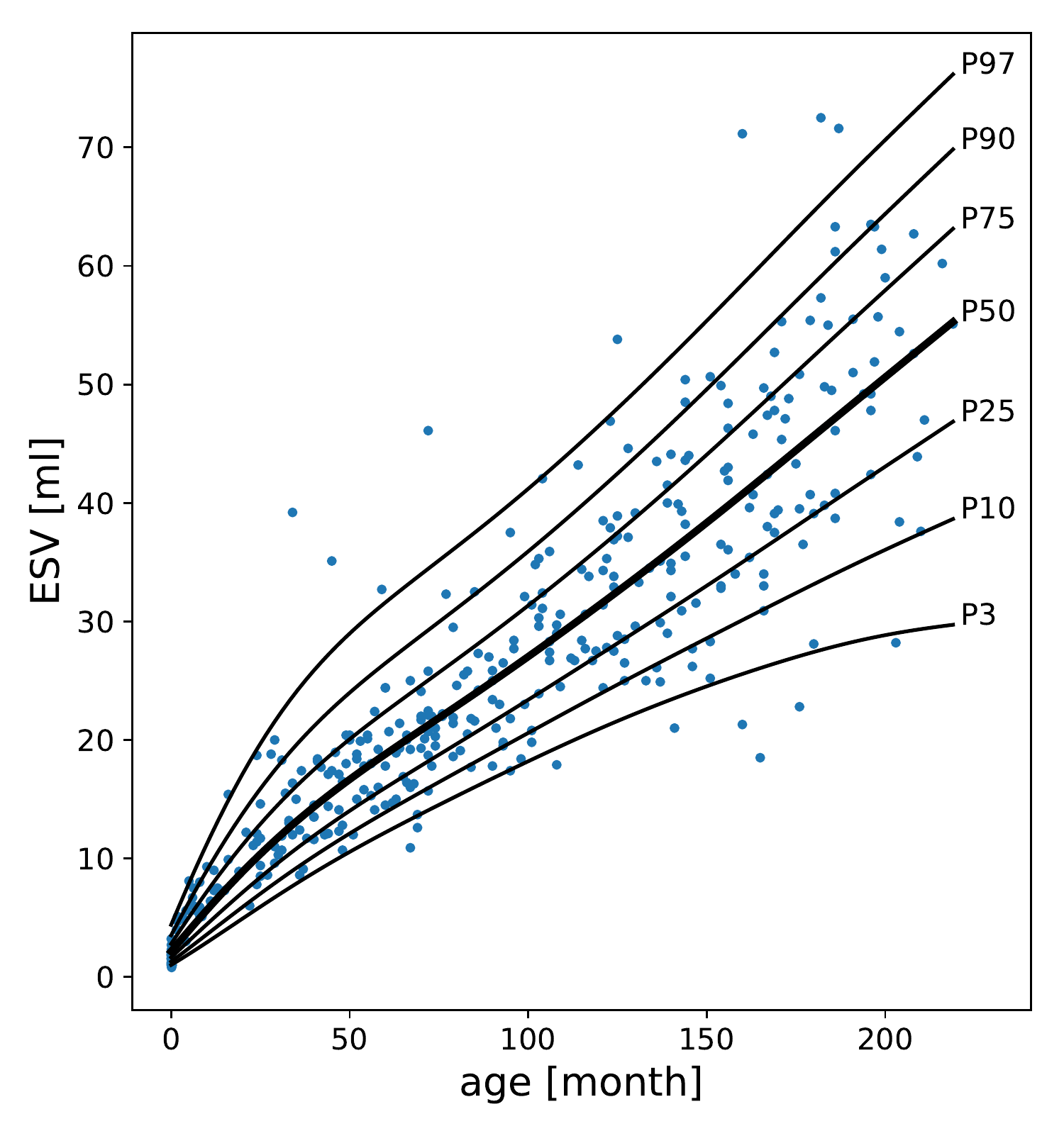}
\subcaption{\scriptsize\code{L\_df} = 1, \code{M\_df} = 2, \code{S\_df} = 1}
\end{subfigure}
\begin{subfigure}[c]{.3\textwidth}
\includegraphics[width=1\textwidth]{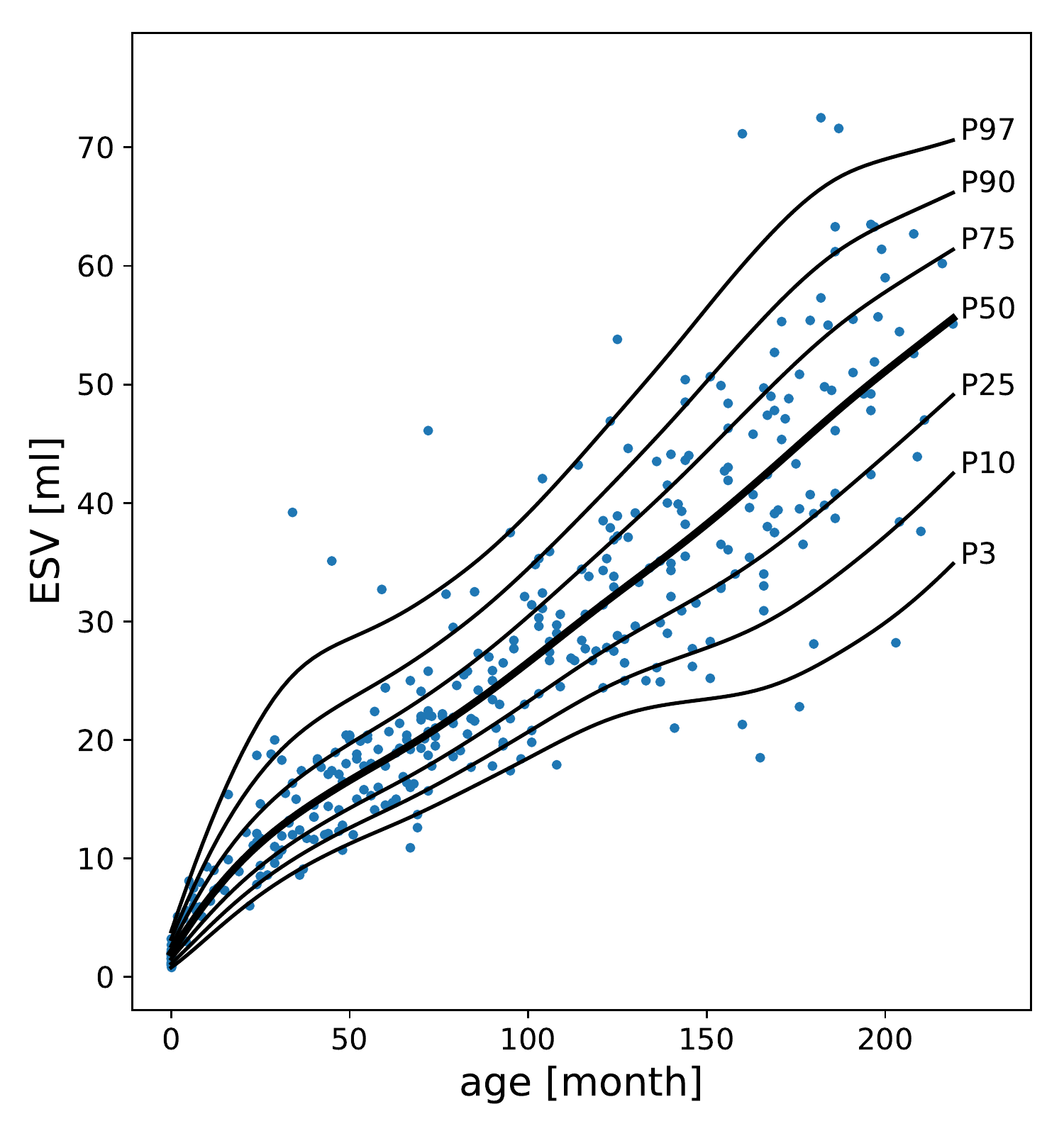}
\subcaption{\scriptsize\code{L\_df} = 4, \code{M\_df} = 4, \code{S\_df} = 4}
\end{subfigure}

\caption{\label{fig:modelFit} \textbf{Model fitting with different settings for the hyperparameters, \code{L\_df}, \code{M\_df} and \code{S\_df}.}}
\end{figure}

For model fitting, the selected data is passed as training dataset to the \pkg{gamlss} function. Users can specify the hyperparameters in the model fitting window ("Model" $\rightarrow$ "Model Fitting"). The hyperparameters for the LMS method are the degree of freedom (\code{df}) for the penalized splines of L, M, and S. After the fitting, a text output field provides a summary of the fitting results.
We recommend a value for \code{df} between 0 and 5 respectively. The effect of different settings for the hyperparameters, \code{L_df}, \code{M_df} and \code{S_df}, on the resulting percentile curves is shown in Figure \ref{fig:modelFit}. The higher the value for \code{df} of the penalized spline is, the higher the flexibility of the curves will be. 

The plot viewer depicts resulting percentile curves after the computation. The text output shows the output of the  \code{gamlss()} function. The output gives information about the fitting results and diagnostic values such as the global deviance.
\subsubsection{Advanced model fitting}
In the advanced model fitting ("Model" $\rightarrow$ "Model Fitting (advanced)"), GAMLSS model settings can be customized. Table \ref{tab:gamlss} shows a list of distributions and smoothing functions for GAMLSS models.

\begin{table}[h]
\centering
\begin{tabular}{|l||l|}
\hline 
Distribution & \proglang{R} function  \\ 
\hline 
\hline
Box-Cox Cole and Green & \code{BCCG()} \\ 
\hline 
Box-Cox power exponential & \code{BCPE()} \\  
\hline 
Box-Cox-t & \code{BCT()} \\  
\hline 
\end{tabular}
\quad
\begin{tabular}{|l||l|}
\hline 
Smoothing function & \proglang{R} function  \\ 
\hline 
\hline
Cubic splines & \code{cs()} \\ 
\hline 
Polynomials & \code{poly()} \\ 
\hline 
Penalized splines & \code{pb()} \\ 
\hline 
\end{tabular}
\caption{\textbf{Distributions and smoothing functions for GAMLSS models.}} 
\label{tab:gamlss}
\end{table}

A full list with distributions and smoothing functions are presented in \cite{gamlss2007}.\\

An example GAMLSS model with a BCCG distribution and cubic splines as smoothing function is: 

\begin{CodeChunk}
\begin{CodeInput}
GAMLSS_model <- gamlss(y ~ cs(x, df = 1),
		       sigma.formula = ~ cs(x, df = 0)),
		       nu.formula = ~ cs(x, df = 0),
		       family = "BCCG",
		       method = RS(),
		       data = dataset_training)
\end{CodeInput}
\end{CodeChunk}

\begin{figure}[htbp]
\centering
\includegraphics[width=0.8\textwidth]{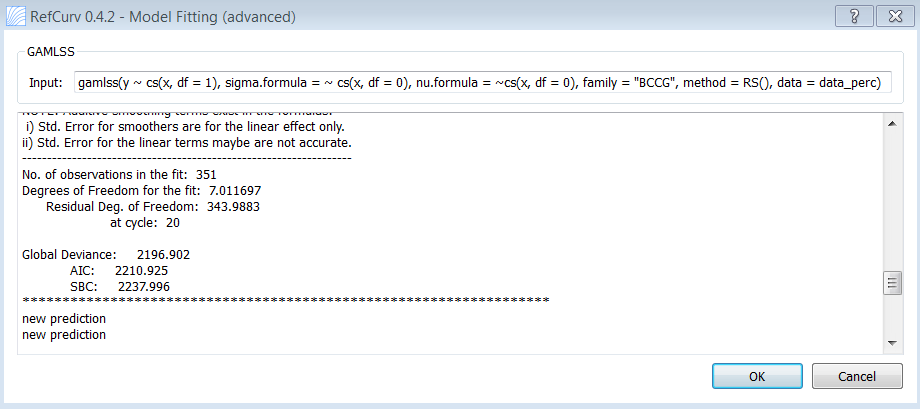}
\caption{\label{fig:refcurv_model_fit_ad}\textbf{RefCurv's advanced model fitting.} }
\end{figure}

In RefCurv, the model fitting with this setting can be realized by typing the command to the input text field of the advanced model fitting window (figure \ref{fig:refcurv_model_fit_ad}).

The features model selection and sensitivity analysis are only available for LMS models. GAMLSS models with other settings are so far not supported.

\newpage
\subsubsection{Outlier detection}
Outliers in the training dataset might have an adverse effect on the model fitting. Datasets might contain outliers because of transcription errors, for instance. RefCurv offers a fast way to detect and analyze potential outliers. Users can decide to exclude individual outliers consequently. 
\begin{figure}[htbp]
\centering
\includegraphics[width=0.9\textwidth]{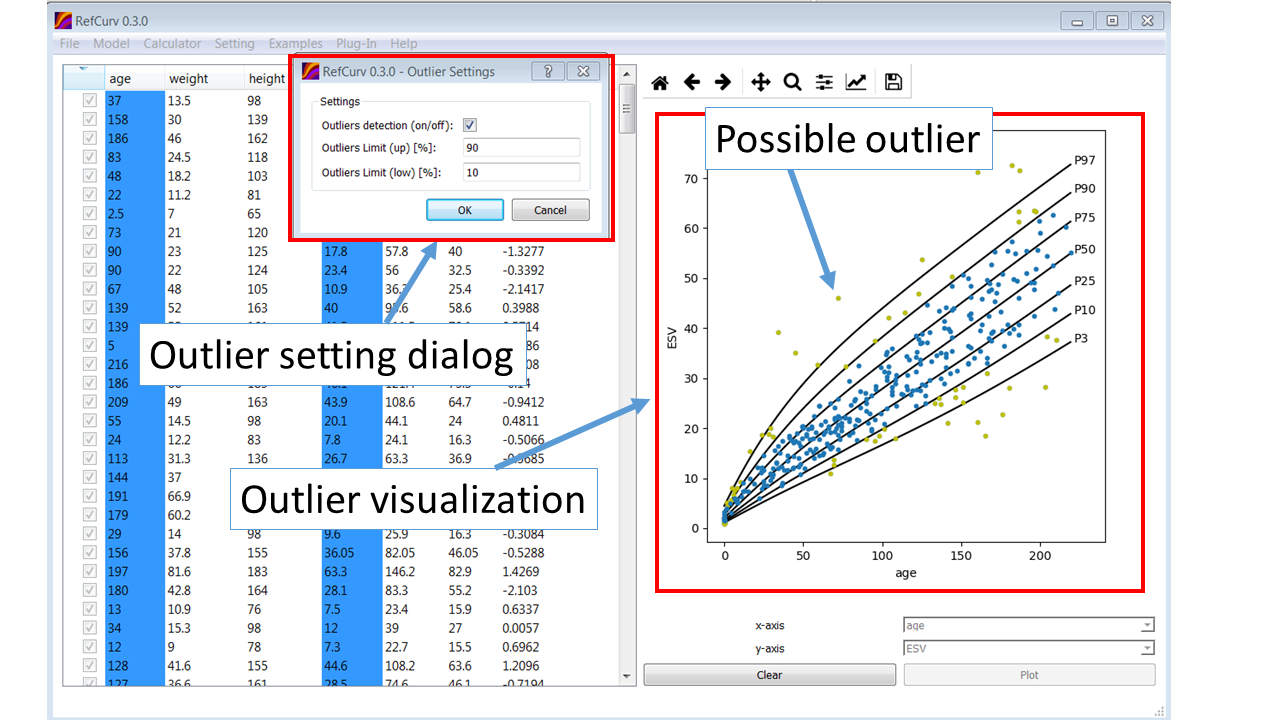}
\caption{\label{fig:refcurv_outlier} \textbf{RefCurv's outlier detection.}}
\end{figure}
The outlier detection is based on a model fitting result. After a predefined model is fitted, the residuals will be calculated. RefCurv's outlier detection feature allows highlighting data points regarding the residuals. Limits for highlighting data points can be chosen individually. In the example in figure \ref{fig:refcurv_outlier}, we chose to set the limit to the 90\% and 10\% ("Setting" $\rightarrow$ "Outliers Setting"). As a result, data points above the 90th percentile curve and below the 10th percentile curve are highlighted in yellow. Afterwards, users can deselect data points that they consider as outliers. Residuals are added as column in the table so that a quantitative assessment is possible.

\subsubsection{Model selection}
As shown before, the LMS method can have different outcomes depending on the degree of freedom (\code{df}) for the penalized splines of the three parameters L, M and S. The task of the model selection is to find an appropriate setting for the \code{df} and balance the trade-off issue between the goodness of fit and complexity. Overfitting can be avoided with this step. RefCurv provides two different ways of model selection.\\
The first model selection method uses the Bayesian Information Criterion (BIC) as decision support for selection (Appendix \ref{app:bic}). A grid search is performed to find the best model concerning the BIC. The model selection window in RefCurv allows to set the limits for the \code{df} of L, M and S. Default step size is set to 1. The output of the model selection is a list of models ordered by BIC. The \code{df} setting of the model with the lowest BIC is considered as best for the chosen dataset.
\begin{figure}[htbp]
\centering
\includegraphics[width=0.5\textwidth]{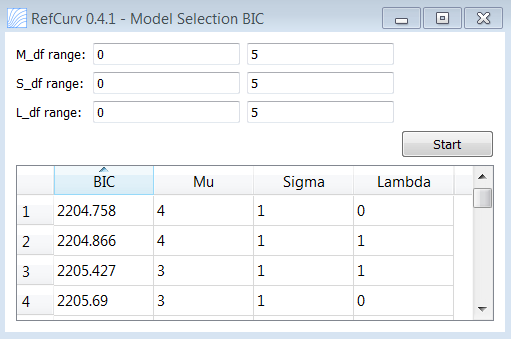}
\caption{\label{fig:refcurv_model_selection} \textbf{RefCurv's model selection.} The range for \code{df} are set to \code{L\_df} = 0,...,5; \code{M\_df} = 0,...,5; \code{S\_df} = 0,...,5}
\end{figure}\\
The second method for model selection is based on cross-validation. RefCurv uses the \pkg{gamlss} function \code{gamlssCV()} for this task. Since datasets are often small in the field of pediatrics, we decided to implement a 10-fold cross-validation. The validation is performed on the training dataset. For that, the dataset is split into ten folds. As a next step, the model is trained on nine folds of the dataset, while the remaining part serves as a validation dataset. Afterwards, the global deviance for the validation dataset is computed, which gives information about the generalization error of the model. Stepwise, each of the ten folds has served as a validation dataset. Finally, the overall generalization error is computed as the mean of the global deviances.\\ 
A cross-validation can be time-consuming due to its computational effort. The model selection based on the BIC is faster and therefore computationally more efficient. Furthermore, RefCurv's BIC method is automatized in the form of a grid search. For the practical application, we currently recommend the BIC method as the model selection for users with little statistical background knowledge.
\subsubsection{Sensitivity analysis}
\begin{figure}[htbp]
\centering
\includegraphics[width=0.7\textwidth]{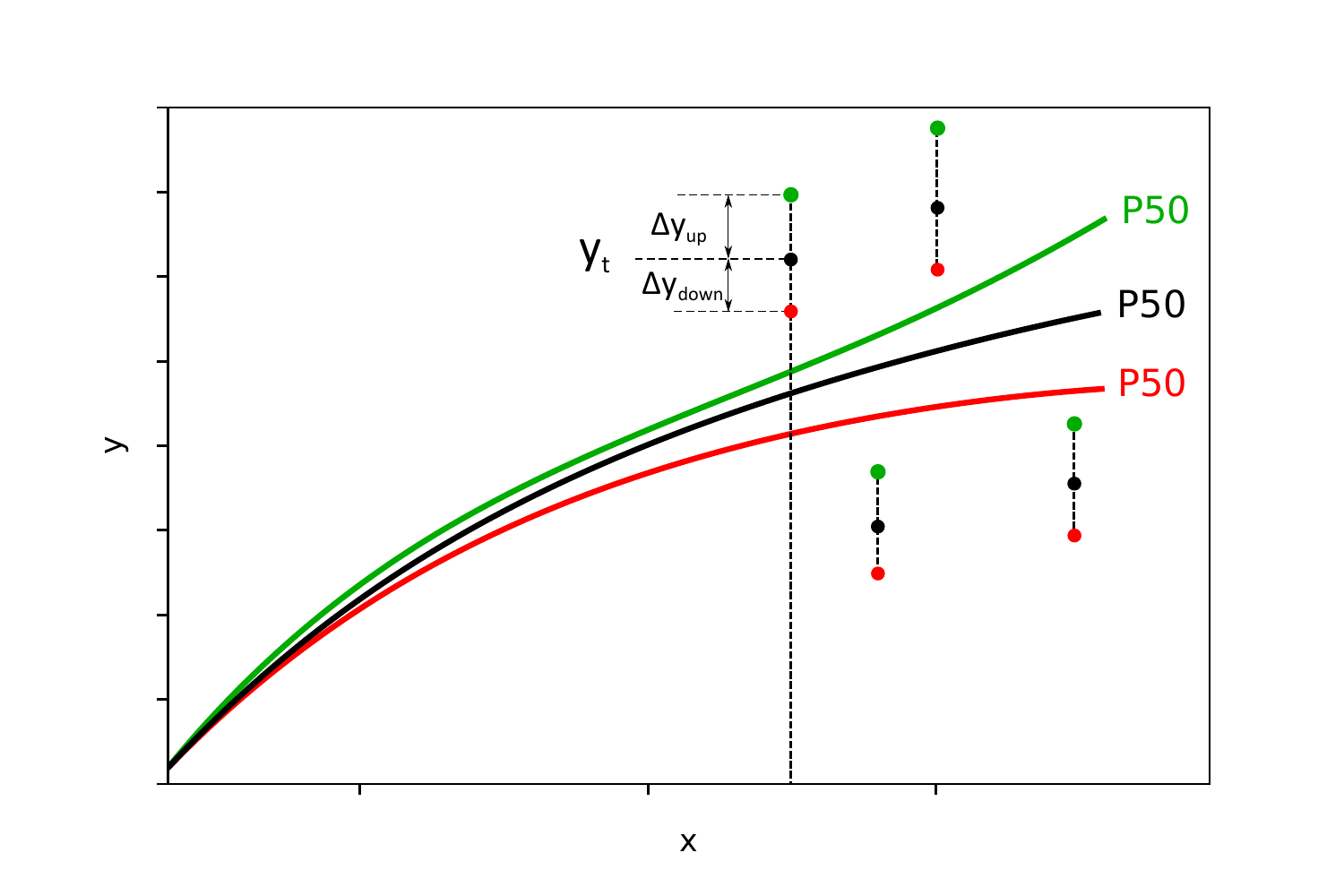}
\caption{\label{fig:sensitiv} \textbf{Concept of the sensitivity analysis.}  A model is fitted to each of the training datasets (red, black, green), which symbolically consists of four data points in this figure. The 50th percentile curve is shown for each of the three cases in the corresponding color.}
\end{figure}

In order to analyze the sensitivity of the fitting method, RefCurv offers a feature to add noise to data points. This kind of uncertainty could be caused by measurement errors. Figure \ref{fig:sensitiv} shows the concept of the sensitivity analysis.
\\
Users can choose single or multiple data points, which are depicted in black. The variations $\Delta y_{\text{up}}$ and $\Delta y_{\text{down}}$ can be applied concerning the chosen response variable $y$. As a result, there are three different datasets (black, green, red), which will be used as training data. The method then fits a model and shows the percentile curves for each of the three cases. In figure \ref{fig:sensitiv} the 50th percentile curve is depicted (black, green, red). \\

\begin{figure}[htbp]
\centering
\includegraphics[width=0.9\textwidth]{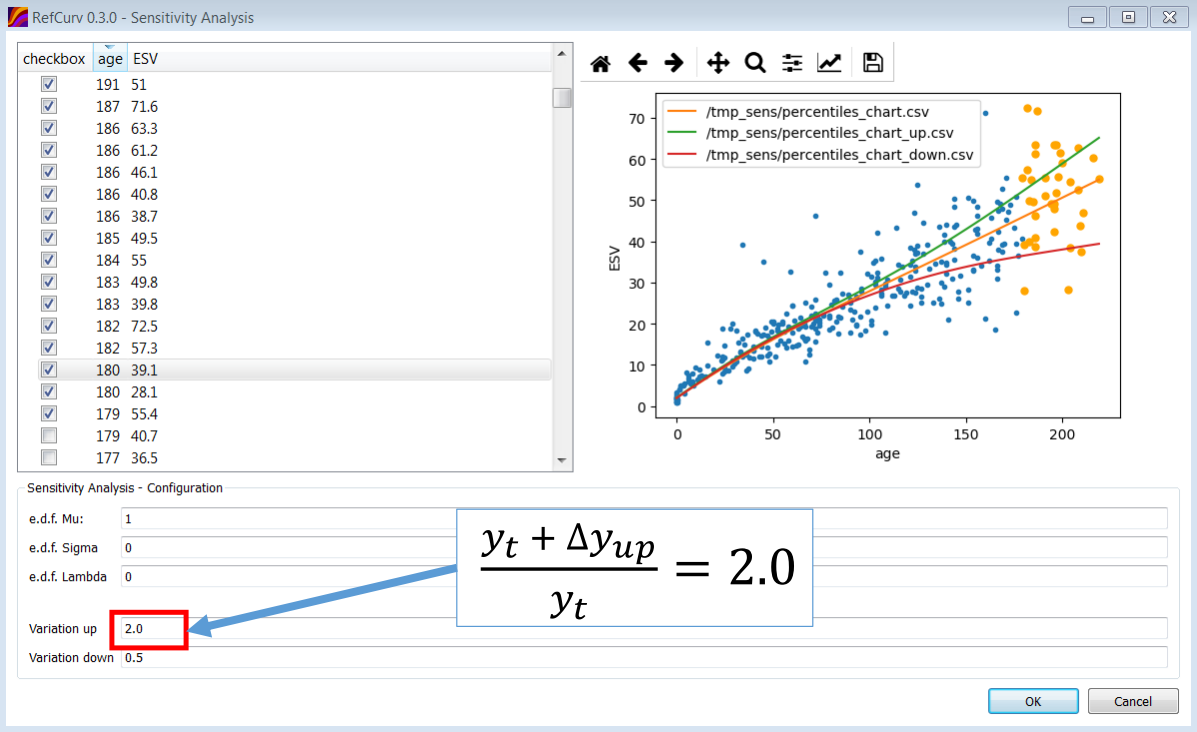}
\caption{\label{fig:refcurv_sensitivity_analysis} \textbf{RefCurv's sensitivity analysis.} Chosen data points with variation are highlighted in yellow. The curves represent the 50th percentile curve for the three cases: variation up, variation down and no variation.}
\end{figure}

Figure \ref{fig:refcurv_sensitivity_analysis} shows an example of the sensitivity analysis in RefCurv. Chosen data points with variation are highlighted in yellow. The values for the variation can be set by the user in the text fields below.\\
This feature also allows to examine the influence of data points on the percentile curves. By varying one data point, for example, we can analyze the effect on the 50th percentile curve.

\newpage
\subsubsection{Model comparison}
In the model comparison window, users can compare the percentile curves of two models. As an example, models with different settings for the \code{df} can be fitted and compared afterwards to analyze the effect. 

\subsubsection{Reverse computation}
For decision support, clinicians often use reference curves or charts from the literature. One problem is that the distribution parameters (L, M, and S for the BCCG distribution) are often missing. RefCurv's reverse computation feature enables users to approximate L, M, and S values for given reference curves. With this method, it is possible to express any reference curve as a LMS model. To achieve that, the method fits a BCCG distribution to the reference curves. The results are the distribution parameters L, M, and S for each value of the covariate. More mathematical details about this approach is given in the Appendix \ref{app:revcomp}.

\subsubsection{Export}
Resulting reference curves can be exported ("File" $\rightarrow$ "Save Reference Curves") as a graph (all common graphical formats) or as a table (CSV file). The values for L, M, and S are automatically exported so that percentiles or z-scores can be computed manually. For clinical use, the values for L, M, and S are essential to compute the z-score of a new case using Cole's formula (Appendix \ref{app:lms}).

\subsubsection{Z-Score/Percentile converter}
\begin{figure}[htbp]
\centering
\includegraphics[width=0.4\textwidth]{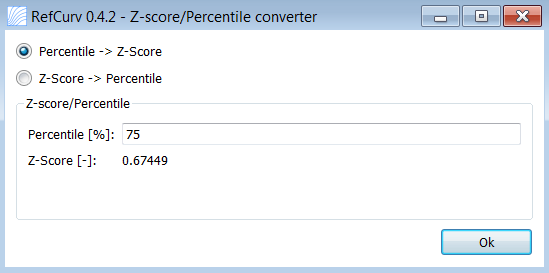}
\caption{\label{fig:refcurv_zscore_converter} \textbf{RefCurv's Z-score/Percentile converter} }
\end{figure}

Percentiles can be converted into z-scores and vice versa. It depends on the examination which of both terms is used by the clinician. RefCurv offers a converter to deal with both definitions ("Calculator" $\rightarrow$ "Z-score/Percentile Converter"). In figure \ref{fig:refcurv_zscore_converter}, we converted the percentile value of 75 to a z-score. In that case, we receive a z-score of 0.67449 as the result.

\subsubsection{Z-Score calculator}
\begin{figure}[htbp]
\centering
\includegraphics[width=0.4\textwidth]{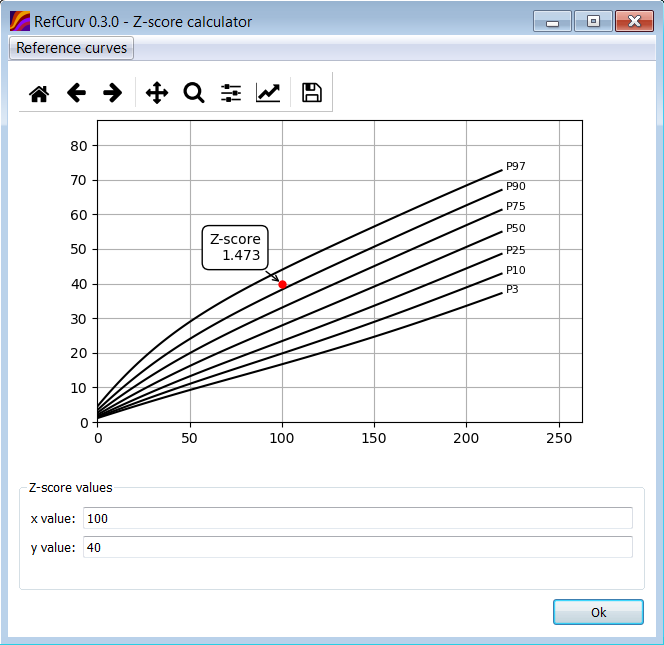}
\caption{\label{fig:refcurv_zscore_calculator} \textbf{RefCurv's z-score calculator.} The z-score of the entered data point (x = 100, y = 40) is 1.473.}
\end{figure}

Clinicians obtain percentile and z-score values of patients as a diagnostic parameter. These values have to be computed with measured data and for a given reference curve. Currently, there is a big number of web and smartphone applications to compute the z-score. With RefCurv's z-score calculator ("Calculator" $\rightarrow$ "Z-score Calculator"), it is possible to compute z-score values of patients directly after the construction of reference curves. \\
Figure \ref{fig:refcurv_zscore_calculator} shows an example where the z-score for the entered data point is 1.473.

\subsubsection{Monte Carlo Simulation}
RefCurv's Monte Carlo Simulation is a feature that could help researchers to design a study. The goal is to plan the required sample size for the construction of reference curves. The simulation is based on a GAMLSS model to create a random sample. Users can enter the simulated sample size in this step. \\
Next, this simulated random sample can be used for the model fitting and analyzing with different settings. Based on this approach, users gather information about the behavior of the models fitted to the created sample size. As a result, users might estimate an appropriate sample size for the construction of reference curves.
\newpage
\subsection{Recommended steps for the modeling of pediatric reference curves with the LMS method} \label{ch24}

The steps for constructing reference curves depend on the analyst's choice. The data analyst could choose the order: model selection, model model fitting, outliers analysis. On the other side, the outlier analysis could also be performed before.\\
Consequently, different approaches can lead to different results and none of these is objective or ideal. However, an unified workflow can improve reliability, comparability and reproducibility. 

\begin{center}
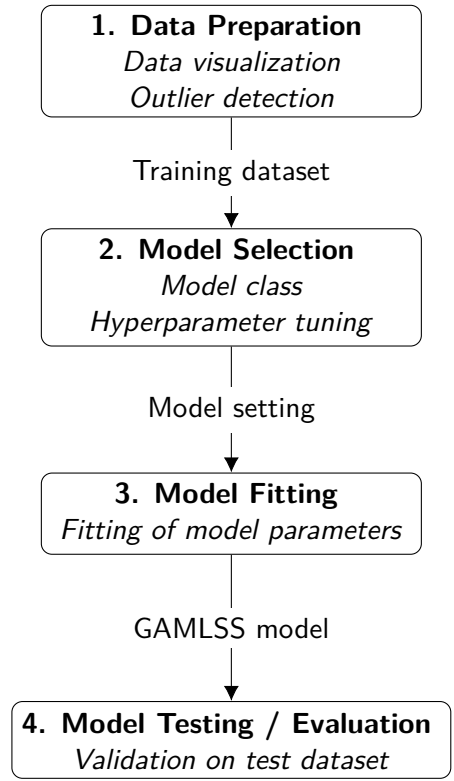

\begin{tikzpicture}[node distance=1.5cm,
    every node/.style={fill=white, font=\sffamily}, align=center]
  \node (dataprep)[standard_style]{\textbf{1. Data Preparation} 
  \vspace{0.3cm} \\ 
  \textit{Data visualization}\\
  \textit{Outlier detection}};
  
  \node (modelsec)[standard_style, below of=dataprep, yshift=-1.5cm] {\textbf{2. Model Selection}
  \vspace{0.3cm}  \\
  \textit{Model class}\\
  \emph{Hyperparameter tuning}};
  \node (modelfit)     [standard_style, below of=modelsec, yshift=-1.5cm]{\textbf{3. Model Fitting}
  \vspace{0.3cm}  \\
  \textit{Fitting of model parameters}};
  \node (modeltest)    [standard_style, below of=modelfit, yshift=-1.5cm]{\textbf{4. Model Testing / Evaluation}
  \vspace{0.3cm}  \\
  \textit{Validation on test dataset}};
   
  \draw[->]     (dataprep) -- node[text width=4cm]
                                   {Training dataset}(modelsec);
  \draw[->]     (modelsec) -- node[text width=4cm]
                                   {Model setting}(modelfit);
  \draw[->]      (modelfit) -- node[text width=4cm]
                                   {GAMLSS model}(modeltest);

\end{tikzpicture}

\captionof{figure}{\label{fig:refcurv_rec_steps}\textbf{Recommended steps for the modeling of pediatric reference curves}}
\end{center}

The \pkg{gamlss} package offers different constellations for the application steps of the LMS method. Steps include model selection and cross-validation. The documentation is very comprehensive (\cite{stasinopoulos2017flexible}). However, we found that \pkg{gamlss} methods are applied in arbitrarily order. A guideline for practitioners seems to be missing. We suggest here steps for the modeling of reference curves with the LMS method, which can be achieved with RefCurv. Figure \ref{fig:refcurv_rec_steps} highlights our four recommended steps.\\

\begin{enumerate}
\item\textbf{Data preparation} is the first step of reference curve modeling. The \textit{data visualization} is crucial to get an overview of the data distribution. We recommend to depict data in a scatter plot and use descriptive statistics to analyze the behavior. The dataset could contain outliers that might have a negative effect on the construction. By highlighting possible outliers, users can reassess, filter and correct the data. We have to make sure that the data serve as a good training set for the model fitting. \\
The output of this modeling step is the training dataset that can be used for the model fitting.
\item As a next step, we recommend to perform \textbf{model selection}. The task of this step is to define the \textit{model class} (distribution family, smoothing functions and hyperparameter), which will be fitted to the data. The decision for the model class should be based on the data distribution, sample size and other data characteristics. Therefore, this step requires experience with modeling.\\
The LMS method uses penalized splines and therefore belongs to the group of non-parametric models. These models can be used if the amount of data is high and a-priori knowledge about the data distribution is missing. In RefCurv, this class is set as default, so that users do not have to deal with complicated model selection tasks.\\
Furthermore, the LMS method contains the \emph{hyperparameters}, \code{df_L}, \code{df_M} and \code{df_S}. These hyperparameters have to be tuned during the model selection.
\item The \textbf{model fitting} follows after the hyperparameters have been found. In this step, the \textit{model parameters} are fitted. For the LMS methods, the model parameters are the vectors L, M and S. The final result of this modeling step is a generalized LMS model that describes the behavior of the data.
\item The last step is the \textbf{model testing / evaluation}. In this step, the model has to be \textit{validated} on an independent test dataset from the population. As a result, users can compute the prediction error for this test dataset, which explains the quality of the model.
\end{enumerate}

\input{application}
\input{discussion}




\section*{Acknowledgments}
\begin{figure}[htbp]
\includegraphics[width=0.5\textwidth]{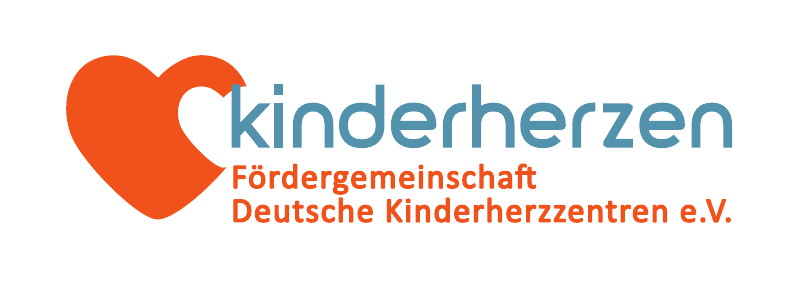}
\end{figure}
This study is funded by Fördergemeinschaft Deutsche Kinderherzzentren e.V.\\

We thank Rupert Hammen and Jochen Kunkel for their help in testing RefCurv.


\newpage
\bibliography{ms}


\newpage
\input{appendix}
\newpage


\end{document}

%% file: application.tex
\newpage
\section[Application]{Application} \label{sec:application}
In this section, we will show how to apply RefCurv on an example dataset, which was acquired in a previous study of our group (\cite{krell2018}). The dataset is accessible for users through the software. First, we will highlight an example where we apply the recommended steps for modeling, which were listed in the previous section. Second, we will go through a case scenario to emphasize the advantages of RefCurv. Last, we will demonstrate how a study design in terms of sample size can be planned.

\subsection{Example} \label{sec:example}
After loading the file ("Examples" $\rightarrow$ "Echo example"), users can observe the data of 351 healthy children in the table viewer. Measured variables are age, weight, height, end-systolic volume (ESV), end-diastolic volume (EDV) and stroke volume (SV) of the left ventricle. The left ventricle is one of the large chambers of the heart and cardiologists measure its volume and shape with echocardiograms. Data from both genders were combined for this example.

\subsubsection{1. Data preparation}
\begin{figure}[htbp]
\centering
\begin{subfigure}[c]{.49\textwidth}
\includegraphics[width=0.95\textwidth]{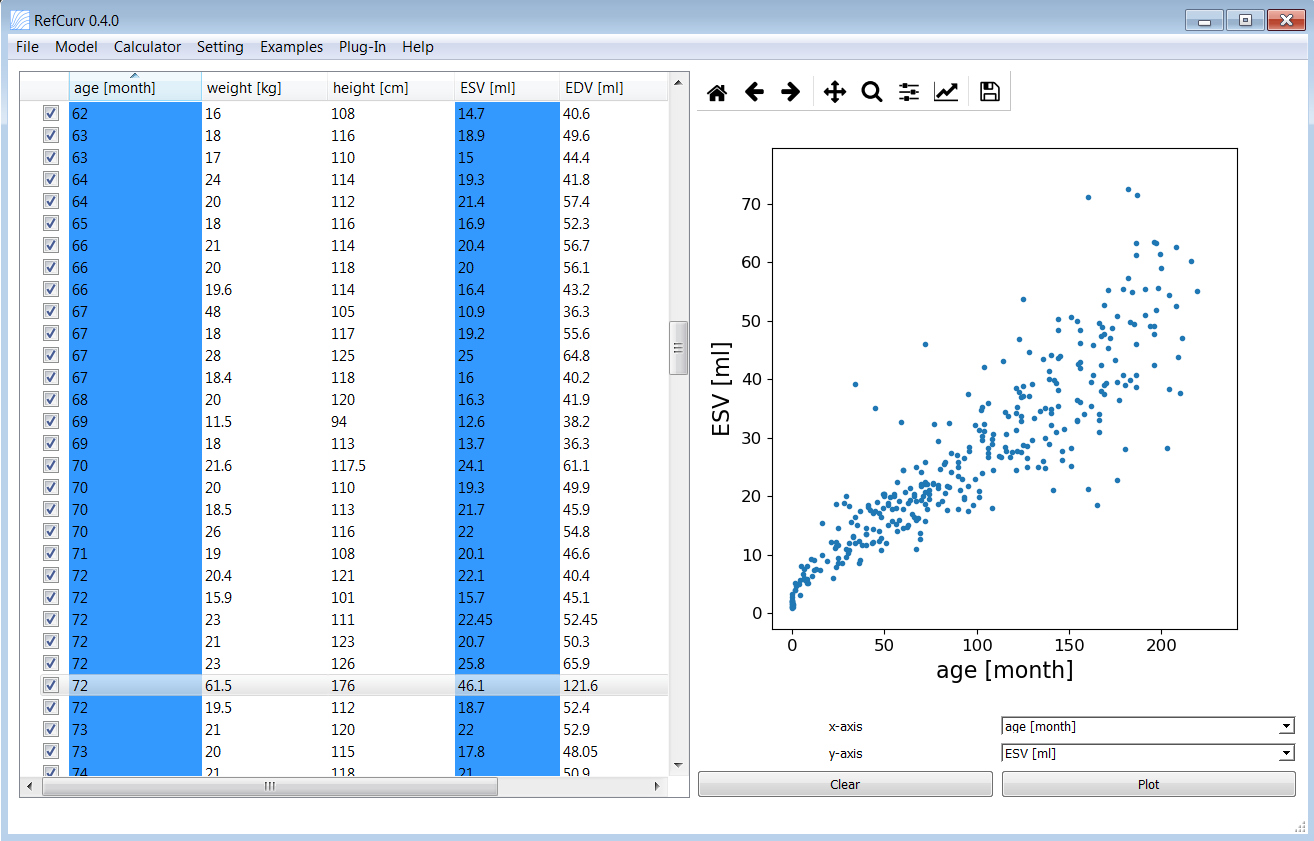}
\subcaption{\scriptsize Data visualization.}
\end{subfigure}
\begin{subfigure}[c]{.49\textwidth}
\includegraphics[width=0.95\textwidth]{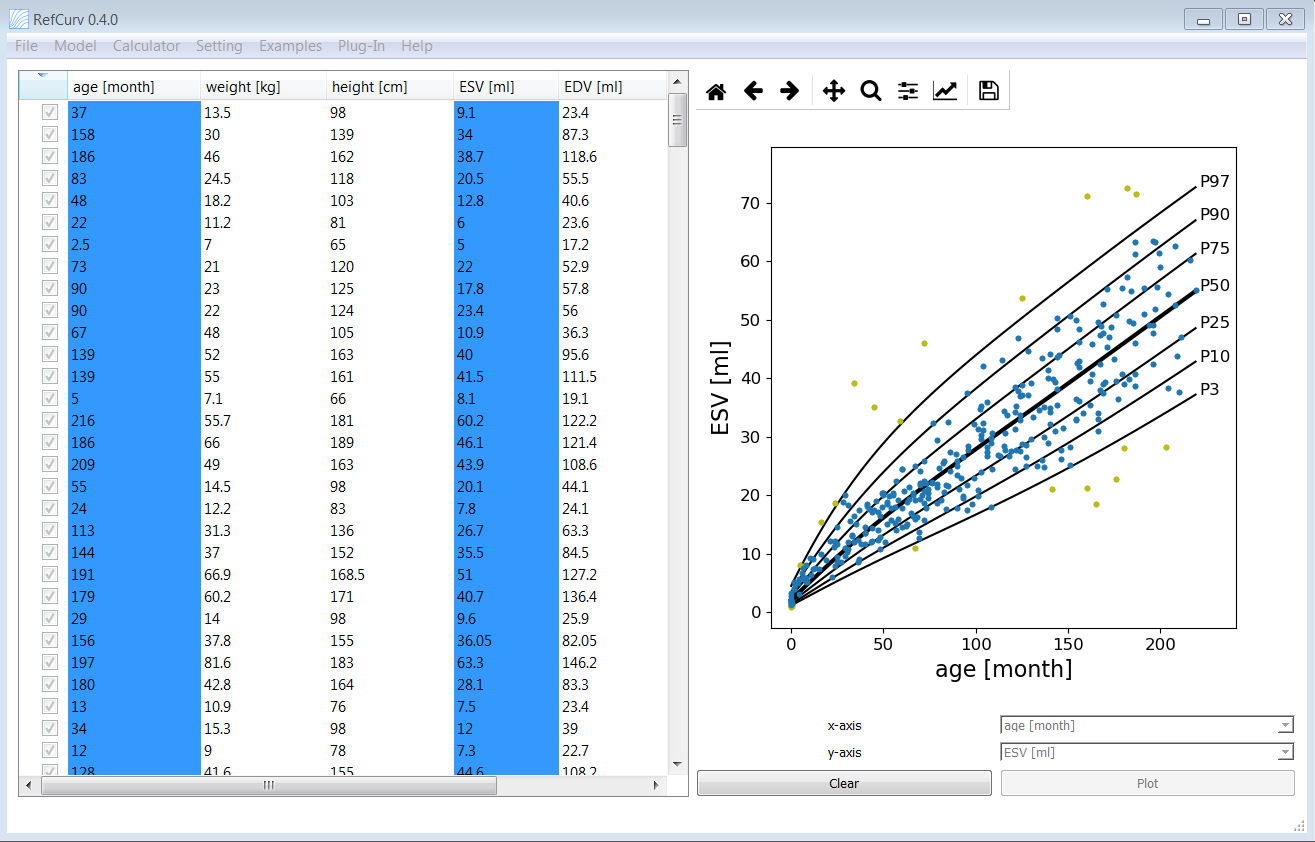}
\subcaption{\scriptsize Outlier detection.}
\end{subfigure}

\caption{\label{fig:example_data_prep} \textbf{Data preparation. Data are visualized (a) and possible outliers are highlighted (b).}}
\end{figure}

First, we examined the data in the table viewer and plotted them as a scatter plot to analyze the data distribution. Age and ESV were selected as variables in the main window. Selecting data points in the table highlights them in the scatter plot (Figure \ref{fig:example_data_prep} (a)).\\
As a next step, we highlighted possible outliers by fitting a standard model (\code{L\_df} = 0, \code{M\_df} = 1, \code{S\_df} = 0) to the data. The limit for highlighting possible outliers was set to the 3rd and the 97th percentile curve. In the interest of simplification, all data below the 3rd percentile curve and above the 97th percentile curve were deselected for this example (Figure \ref{fig:example_data_prep} (b)). Please note that only some of the highlighted data points - the ones that the analyst assesses as abnormal - should be considered as outliers.\\

\subsubsection{2. Model selection}
\begin{figure}[htbp]
\centering
\includegraphics[width=0.4\textwidth]{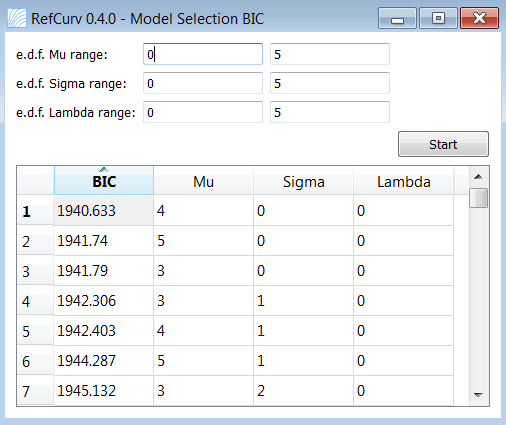}
\caption{\label{fig:example_model_selection} \textbf{Model selection.} The range for \code{df} are set to \code{L\_df} = 0,...,5; \code{M\_df} = 0,...,5; \code{S\_df} = 0,...,5}
\end{figure}

In order to optimize the hyperparameters \code{L\_df}, \code{M\_df}, and \code{S\_df}, RefCurv's BIC model selection was performed. The range for each \code{df} was set to be 0 to 5.\\
The result of the model selection is shown in Figure \ref{fig:example_model_selection}. The model with setting \code{M\_df}=4, \code{S\_df}=0 and \code{L\_df}=0 had the lowest BIC (1940.633). This model was chosen as the best model, and its settings were used in the model fitting window to create the new prediction.

\subsubsection{3. Model fitting}
\begin{figure}[htbp]
\centering
\begin{subfigure}[c]{.49\textwidth}
\includegraphics[width=0.99\textwidth]{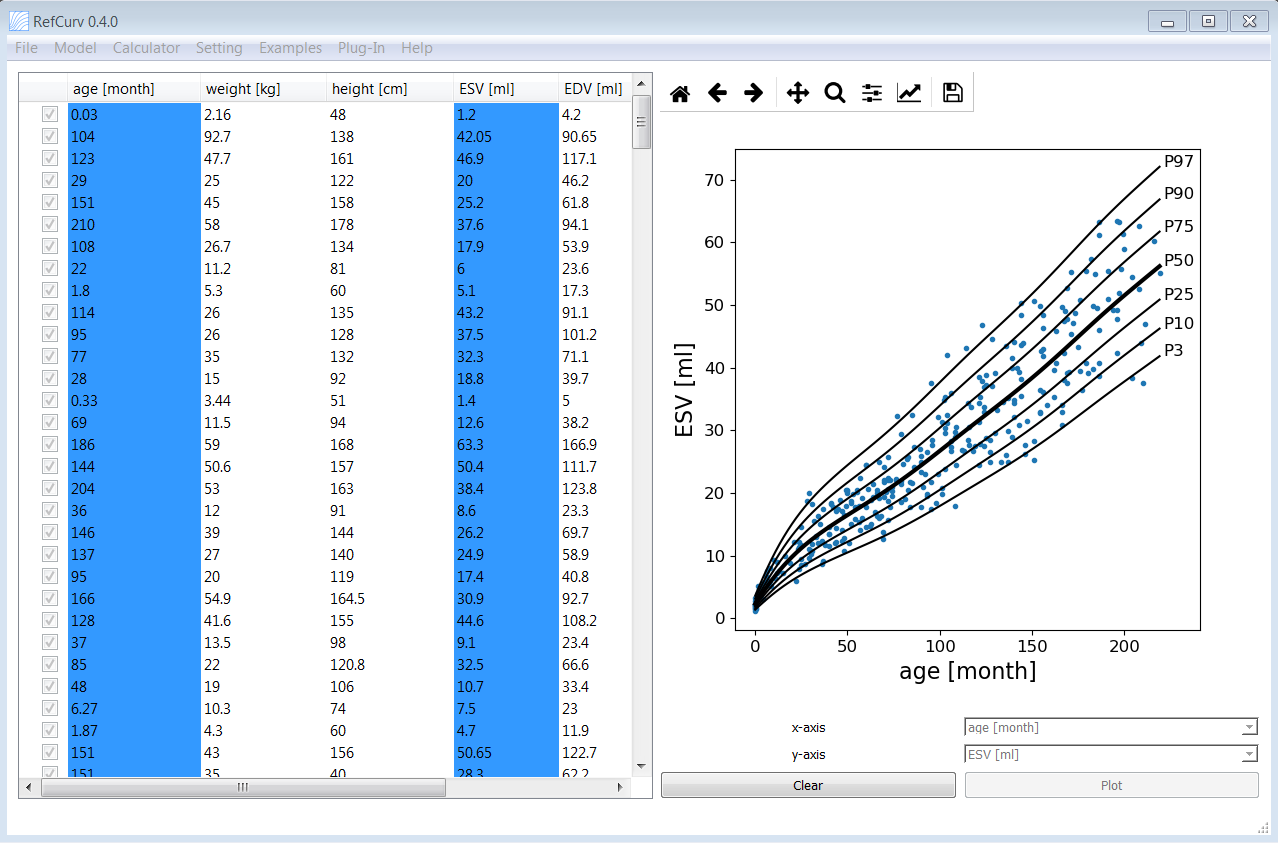}
\end{subfigure}
\begin{subfigure}[c]{.49\textwidth}
\includegraphics[width=0.99\textwidth]{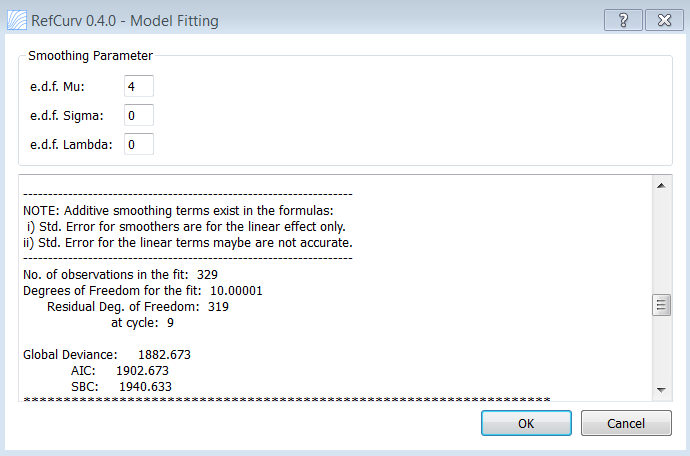}
\end{subfigure}
\caption{\label{fig:example_model_fitting} \textbf{Model fitting.} The \code{df} are set to \code{M\_df}=4, \code{S\_df}=0, and \code{L\_df}=0}
\end{figure}

The model was fitted with the tuned hyperparameters (\code{M\_df}=4, \code{S\_df}=0, \code{L\_df}=0). \\
Figure \ref{fig:example_model_fitting} shows the results of the model fitting process.

\subsubsection{4. Model testing}
\begin{figure}[htbp]
\centering
\includegraphics[width=0.6\textwidth]{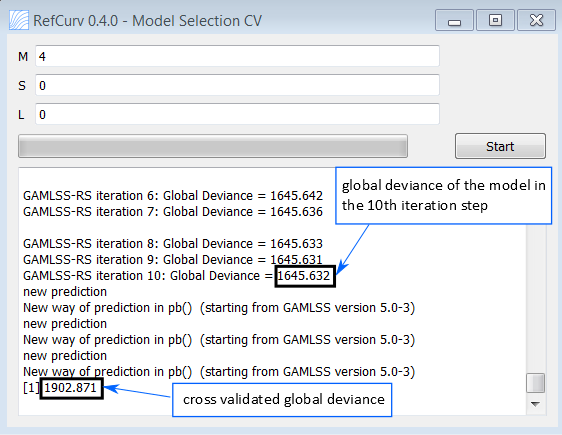}
\caption{\label{fig:example_model_testing} \textbf{Model testing.} We used a 10-fold cross-validation to determine the cross validated global deviance of 1902.871 for the model. The global deviance for the training of the last model during the cross validation (10th iteration step) was 1645.632.}
\end{figure}

As the last step, the model was validated by using the implemented 10-fold cross-validation function. In the model validation window, the LMS-values, which were found through the model selection process, were given (Figure \ref{fig:example_model_testing}). The cross-validated global deviance was 1902.871 for this case.

\newpage
\subsection{Case scenario}
In order to display the other features of RefCurv, we present a case scenario for the construction of reference curves. First, we studied the impact of reducing data points and creating a gap in the age range.
We also investigated the effect of the data points on the sides (edges) of the measuring range.\\
\begin{figure}[htbp]
\centering
\begin{subfigure}[c]{.24\textwidth}
\includegraphics[width=1\textwidth]{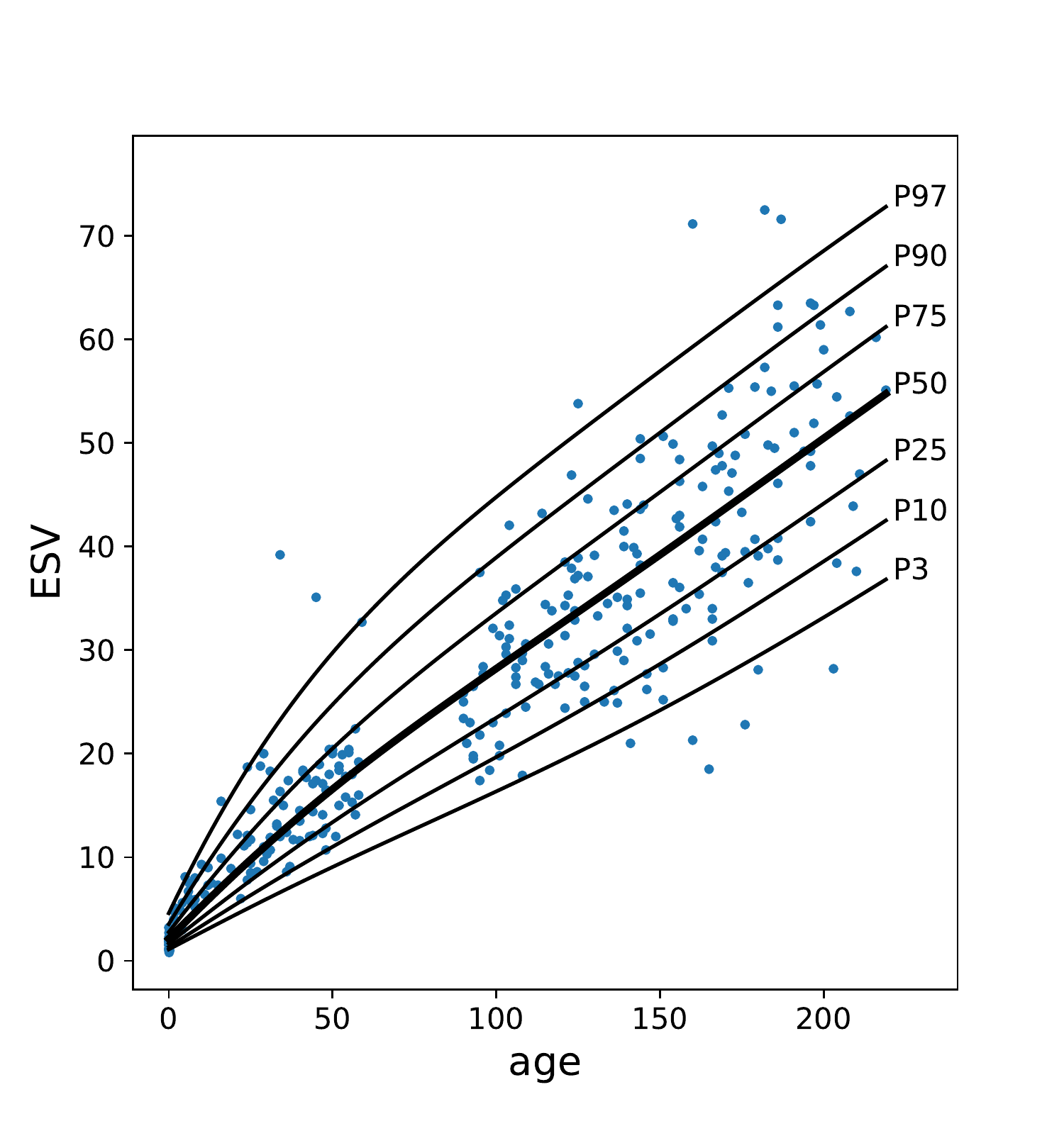}
\end{subfigure}
\begin{subfigure}[c]{.24\textwidth}
\includegraphics[width=1\textwidth]{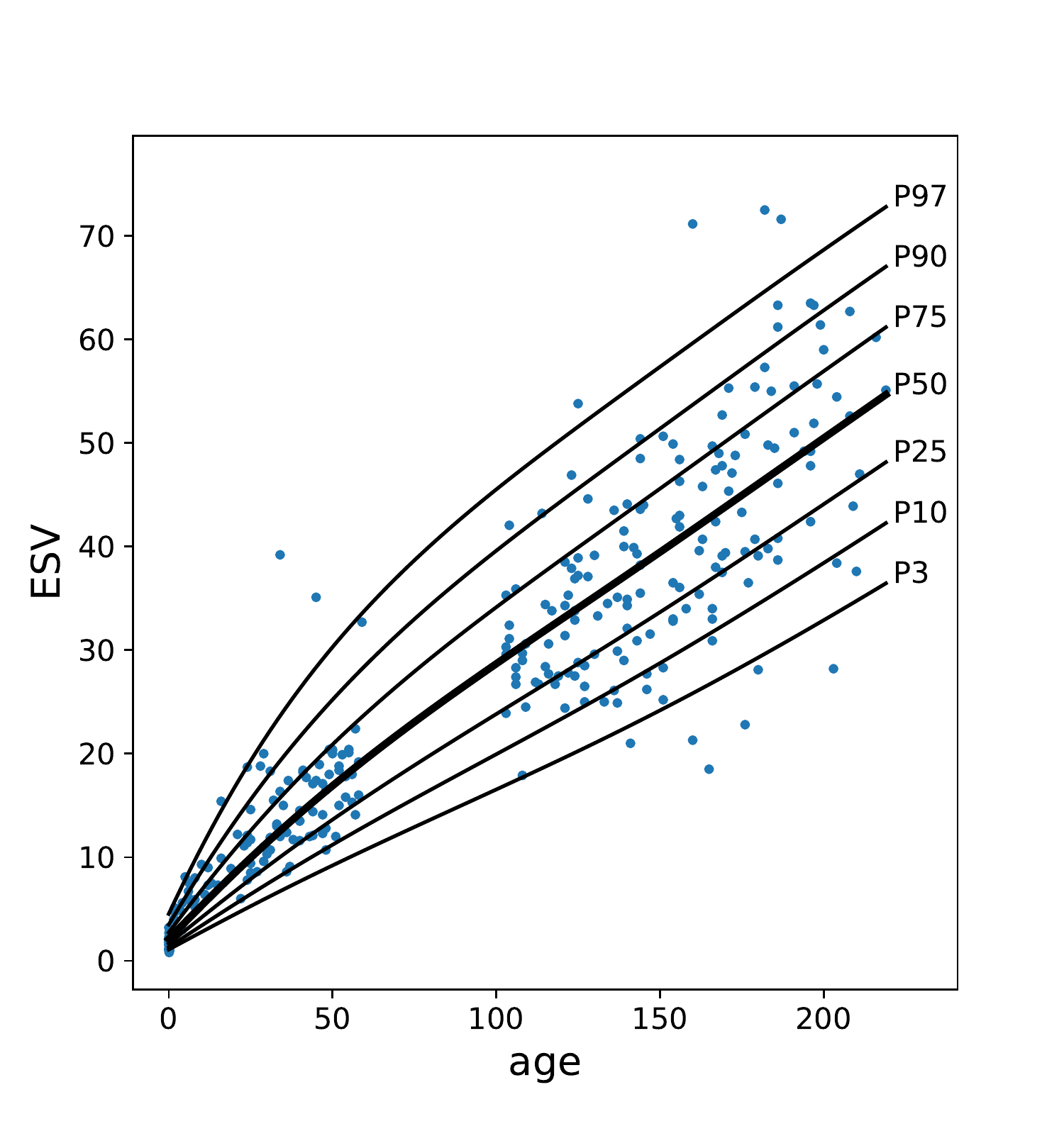}
\end{subfigure}
\begin{subfigure}[c]{.24\textwidth}
\includegraphics[width=1\textwidth]{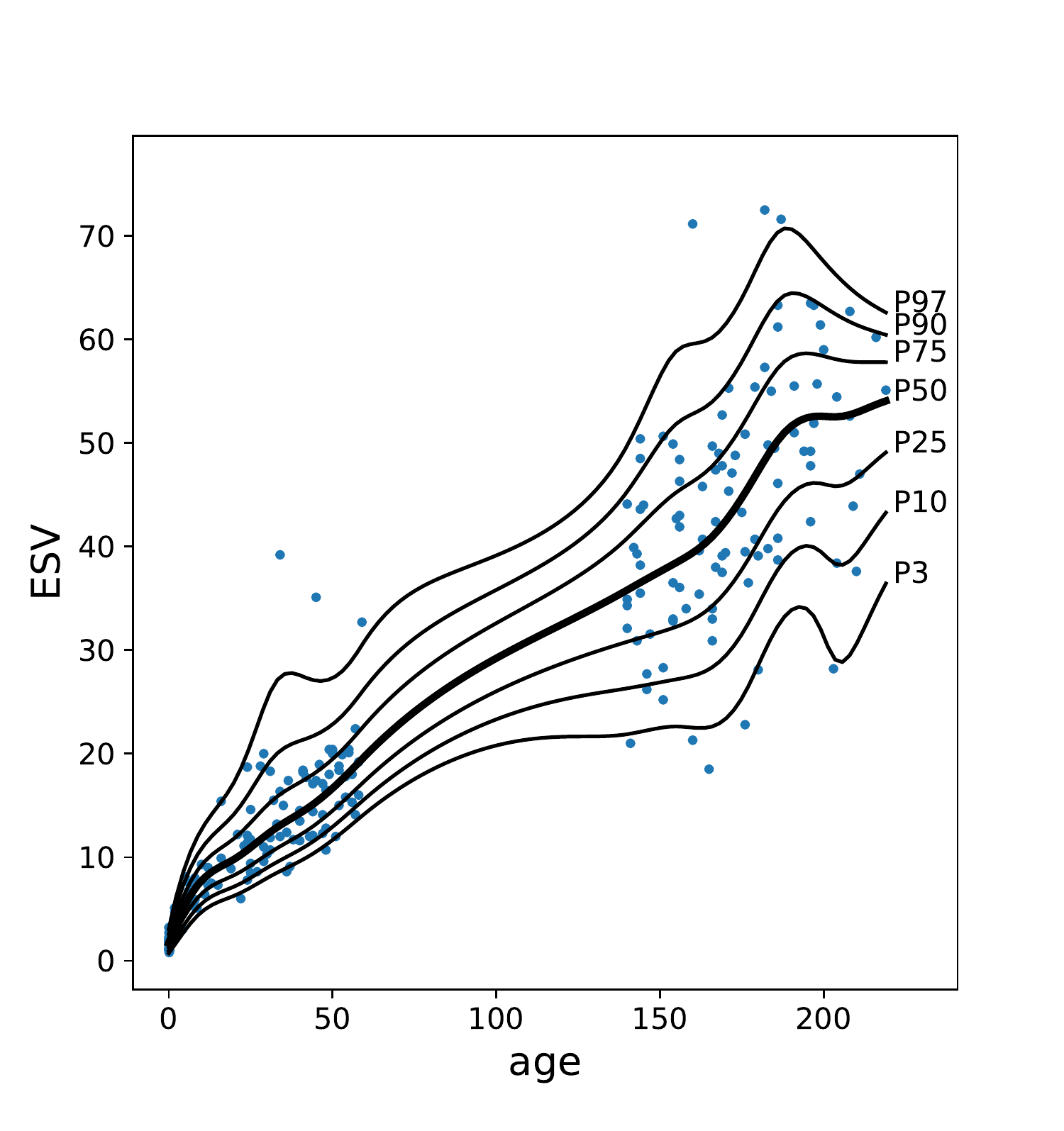}
\end{subfigure}
\begin{subfigure}[c]{.24\textwidth}
\includegraphics[width=1\textwidth]{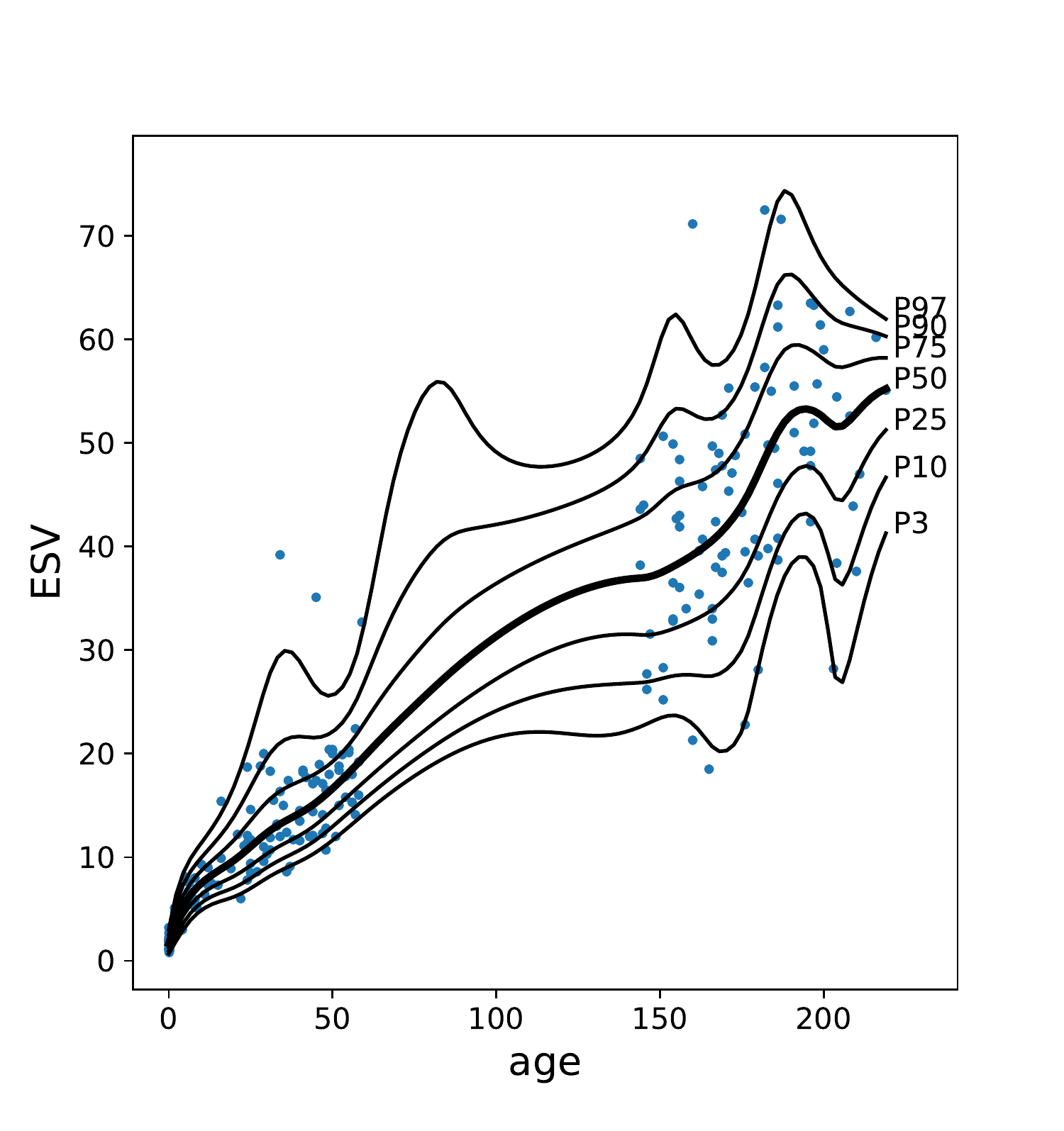}
\end{subfigure}
\caption{\label{fig:case_scenario} \textbf{Case scenario.} We reduced the number of data points (from left to right) creating a gap in the data cloud. }
\end{figure}

\begin{figure}[htbp]
\centering
\begin{subfigure}[c]{.49\textwidth}
\includegraphics[width=1\textwidth]{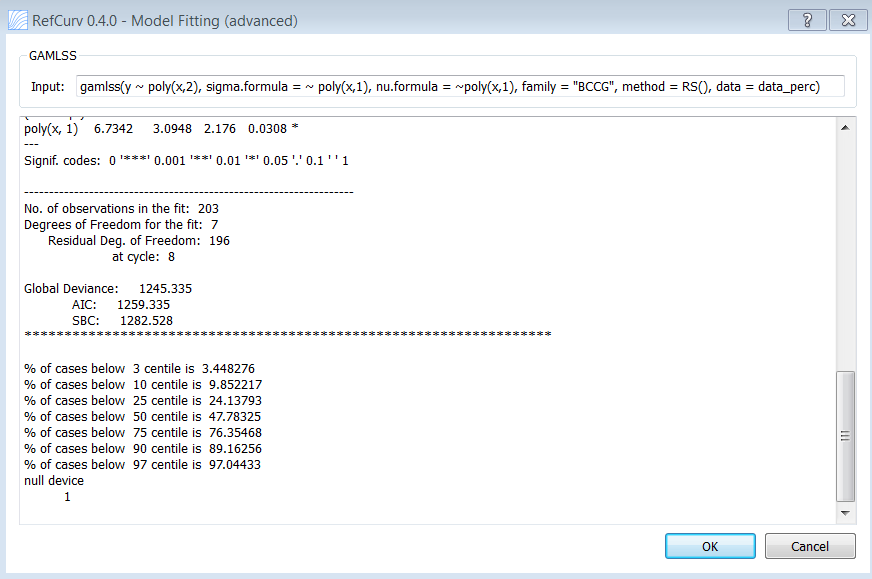}
\end{subfigure}
\begin{subfigure}[c]{.49\textwidth}
\includegraphics[width=1\textwidth]{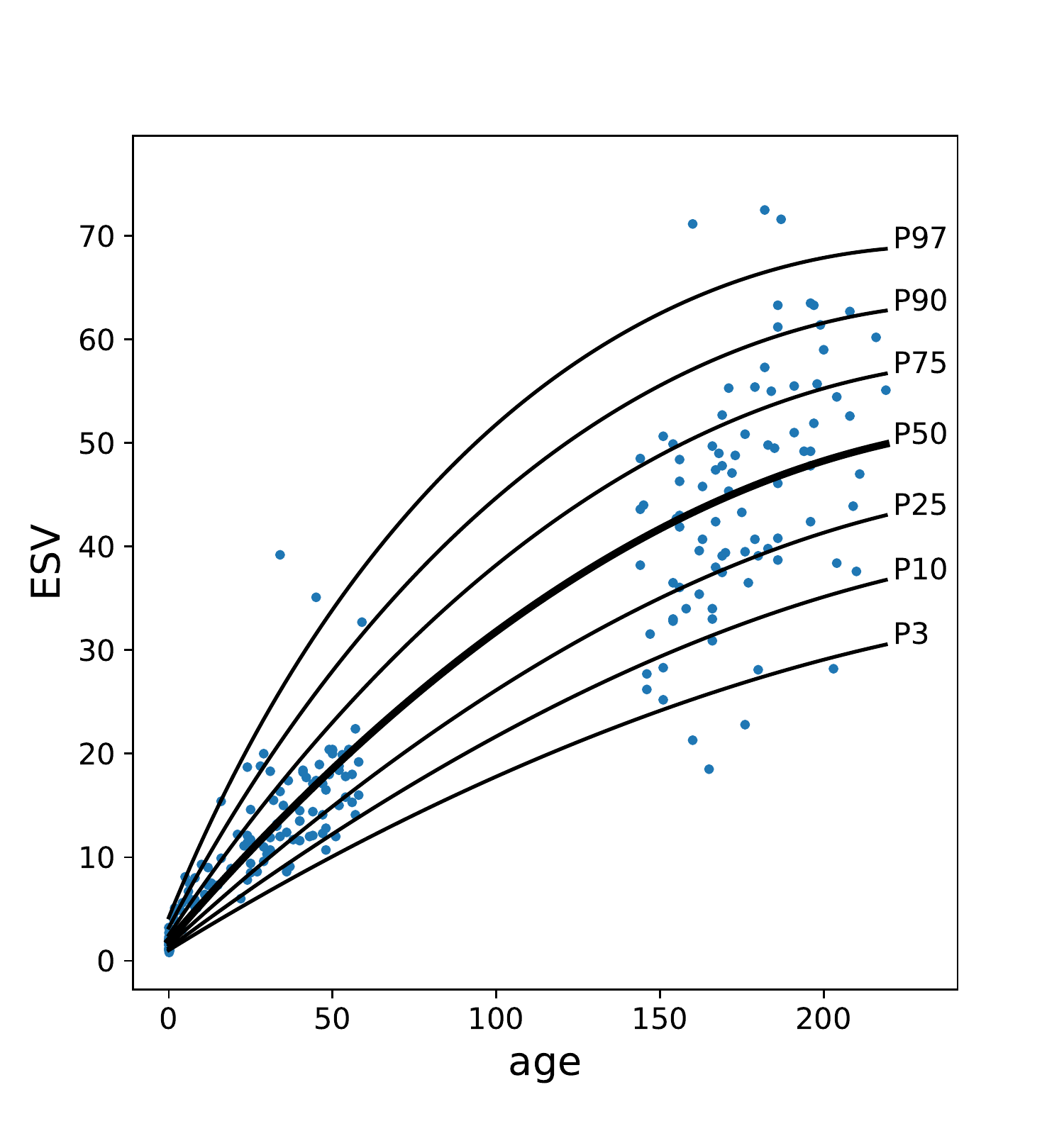}
\end{subfigure}
\caption{\label{fig:advanced_model} \textbf{Advanced model fitting.} We defined a new \pkg{gamlss} model with \code{poly(x)} functions for the curve fitting.}
\end{figure}

In this scenario, data points from the training dataset were gradually excluded in the middle of the data cloud. This resulted in a gap possibly causing computational problems. With this procedure, the feasibility and robustness of the LMS method were tested. Figure \ref{fig:case_scenario} shows the procedure. When the number of data points reached less than 274, the LMS method gave unsatisfying reference curves with low smoothness as a result. A change of the hyperparameters \code{df} did not help to improve the smoothness. \\
To solve this issue, we used the RefCurv's "Advanced model fitting". We defined \code{GAMLSS\_model} with the following setting:

\begin{CodeChunk}
\begin{CodeInput}
GAMLSS_model <- gamlss(y ~ poly(x,2),
		       sigma.formula = ~ poly(x,1),
		       nu.formula = ~ poly(x,1),
		       family = "BCCG",
		       method = RS(),
		       data = dataset_training)
\end{CodeInput}
\end{CodeChunk}
where \code{poly(x)} is the function for evaluating orthogonal polynomials. Figure \ref{fig:advanced_model} shows the result of the fitting.

\newpage
\subsection{Design of study}
\begin{figure}[htbp]
\centering
\begin{subfigure}[c]{.49\textwidth}
\includegraphics[width=0.95\textwidth]{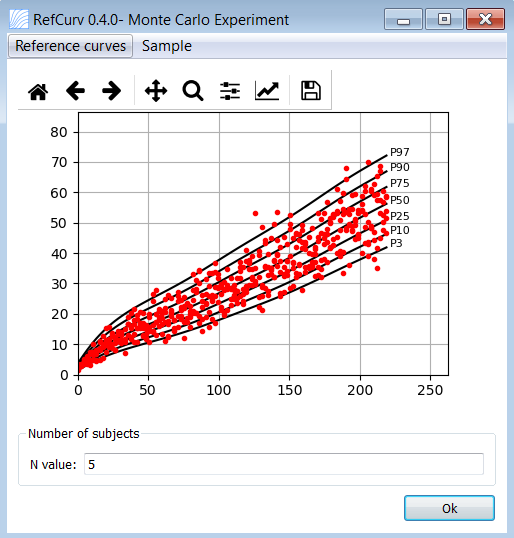}
\subcaption{\scriptsize n = 500}
\end{subfigure}
\begin{subfigure}[c]{.49\textwidth}
\includegraphics[width=0.95\textwidth]{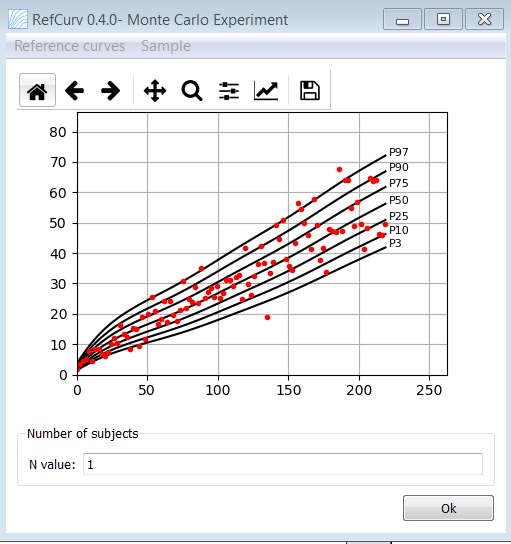}
\subcaption{\scriptsize n = 100}
\end{subfigure}
\caption{\label{fig:design_study} \textbf{Monte Carlo simulation.} Different sample sizes were created from a previously chosen model.}
\end{figure}

RefCurv's Monte Carlo simulation can be used to visualize the impact of this sample size. Let us take a look at the resulting reference curves from the example before (Figure \ref{fig:example_model_fitting}). The curves can be loaded into the simulation window and different sample sizes can be created for the simulation. We chose a sample size of 500 and reduced it to 100 (Figure \ref{fig:design_study}).

\begin{figure}[htbp]
\centering
\includegraphics[width=1\textwidth]{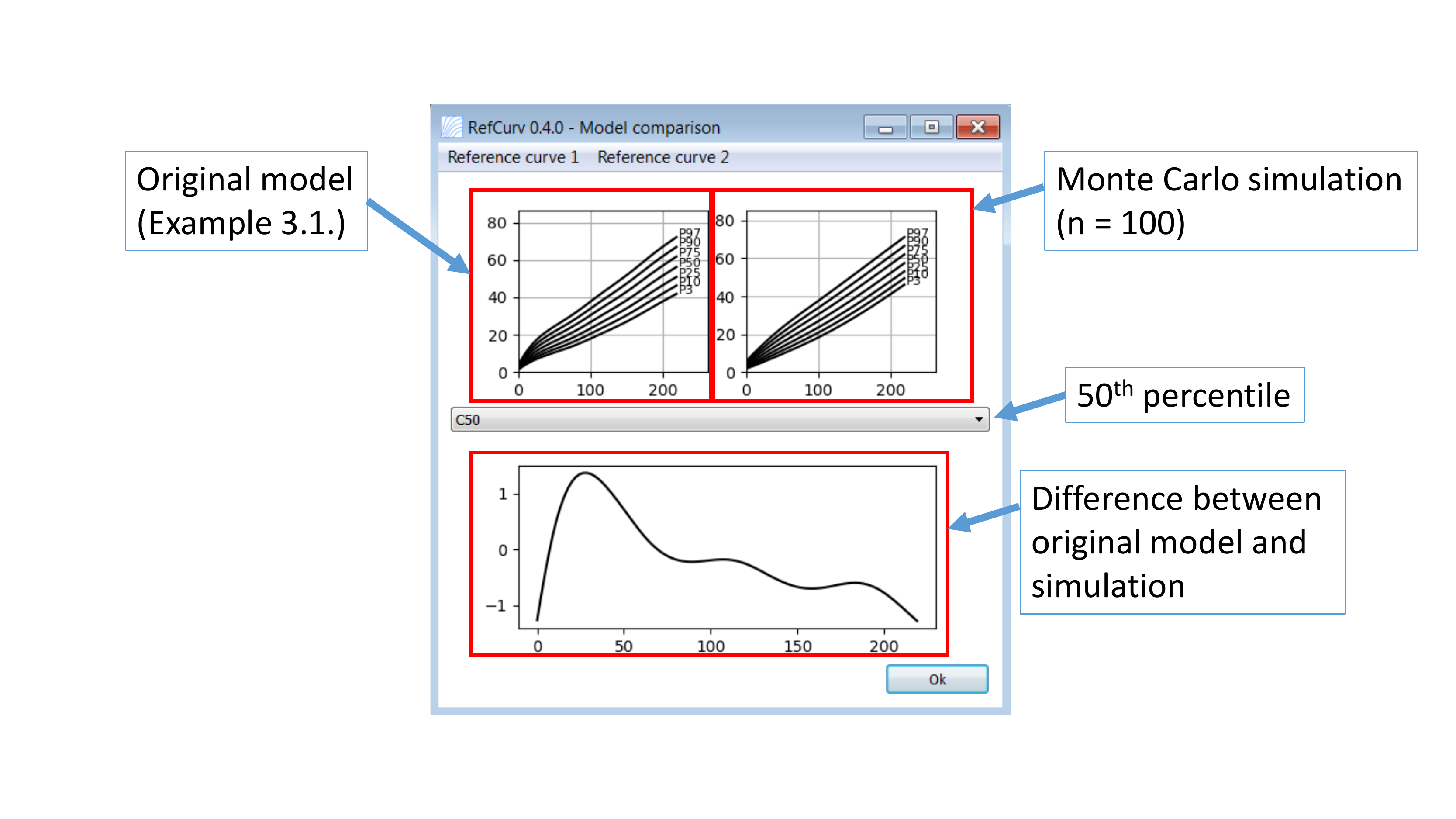}
\caption{\label{fig:model_comparison} \textbf{Model comparison.} We used the model from example \ref{sec:example} and created a sample ($n = 100$) by Monte Carlo simulation. This sample size served as training dataset to fit a model. The 50th percentile of both models is compared.}
\end{figure}

We continued with the lower sample size ($n = 100$), fitted a LMS model with standard setting and compared it to the original model (Figure \ref{fig:model_comparison}). The difference of the 50th percentile curve for both models was compared. It shows that the absolute difference is never bigger than 1 milliliter. From these results, users could conclude that a sample size of 100 might be sufficient to create percentile curves. We recommend to use similar analyses like computing the difference of the other percentile curves to corroborate this assumption.

\newpage
We illustrated that RefCurv's features might help to design studies before data is acquired. This can be achieved by going through a case scenario like the presented one. Questions about the number of data points, data distribution, robustness of the LMS method and impact of outliers can be analyzed in advance. Issues like missing data could be considered during the planning.


%% file: discussion.tex
\section[Discussion]{Discussion} \label{sec:discussion}
In the field of medical research, physicians and life scientists miss an easy-to-use software for the construction of reference curves. For the application of modern statistical approaches, most methods such as the \pkg{gamlss} package are implemented in \proglang{R} and require programming skills. Furthermore, the number of steps for the statistical analysis is high and hampers a quick analysis. Thus, there is a gap between the statistical methods and end-users. To address this issue, we presented RefCurv, a software that enables the construction and analysis of reference curves for children. \\

In this article, we focused on medical and particularly echocardiographical data where reference curves are broadly discussed. \cite{dallaire2011} presented a very detailed analysis of modeling approaches. In their study, they focused on parametric regression models and analyzed features such as goodness of fit for the model and data distribution. A similar approach was presented by \cite{kobayashi2016} while they also added nonparametric regression models to their analysis. These studies have led to the necessity of developing a tool, which can simplify and automate the computation. Our project focused on the LMS method and GAMLSS models because of their good quality and reliability. In this context, the \pkg{gamlss} package provides a broad set of distributions and smoothing functions.\\

With this project, we lay the foundation for further analysis of reference curves. RefCurv helps to solve issues that were discussed in multiple articles before. \cite{cantinotti2018} propose, for example, to develop a uniform approach to data normalization. The same authors developed an application for smartphones, BabyNorm, which enables and simplifies the use of medical reference values in clinics (\cite{cantinotti2017}). The advantage of BabyNorm is the possibility to choose between different published reference curves. Clinicians can compare patient data to normal values, which are given in the journal articles. We found that a quality index for the published reference curves is missing. Users mostly have to choose reference from studies arbitrarily without knowing any details about the references values provided by the study, such as a number of data points, statistical method or goodness of fit. RefCurv can solve this problem by offering methods such as the cross-validation method to rank different models.\\

The development process of RefCurv is ongoing in order to improve the functionality. An automated computation of the sample size is planned. So far, the software was tested with multiple datasets and is found to be stable. However, stability and convergence issues might occur like stated in documentations of the \pkg{gamlss} package (\cite{gamlss2005}). So far, the handling of negative data points has not been considered but will be considered in future versions. In the future, RefCurv will be tested on large and highly distributed data to find out about limitations.\\


RefCurv can help to standardize procedures and plan the acquisition of data. The design of the study can be planned in advance through exemplary case scenarios. For example, researchers could simulate the impact of the sample size on their reference curves to find out: (i) the minimum number of data points required, (ii) the effect of an increasing number of data points, (iii) the correct choice of predictor for the curves, (iv) the necessity to stratify by gender or other variables. The construction of percentile curves with RefCurv can determine the impact of these parameters on their study results.\\

A fundamental problem in pediatrics is the low number of measurements because the data acquisition is long lasting, expensive and difficult. Consequently, sample sizes are often small, which is discussed in multiple articles (\cite{tanaka1987big, cantinotti2017, williams2012standard}). A solution for this issue is to acquire data in multicenter studies such as our example dataset (\cite{krell2018}). Merging of data can be easily managed with RefCurv and the effect for different training datasets on the resulting reference curves can be quickly tested. 


\newpage
\subsection{Reuse potential}
We demonstrated the software on echocardiographical data of children. Apart from pediatric applications, reference curves play an important role in many other disciplines of medicine. As an example, they could be used to describe the growth process of organs or the effect of a drug. Beyond that, every natural science or technical environment requires references in order to analyze and improve processes. Thus, RefCurv's application field is flexible.\\

The application of the LMS method have been proven to be valuable for pediatric reference curves. However, model classes and distributions as listed in \cite{gamlss2007} are accessible through the advanced model fitting in RefCurv. Therefore, this software is flexible and can be adjusted according to research hypothesis, theory or purpose.\\


Due to its easy-to-use GUI, the application does not need any extended training but can be applied quickly. After the installation, data visualization, model fitting and reference curve analysis are intuitive.\\

RefCurv uses GAMLSS models in a \proglang{Python} environment, which opens the door for combinations with other \proglang{Python} software packages. The Simvascular project (\cite{simvascular}), for example, offers methods to model the cardiovascular system and provides a \proglang{Python} interface. Computation results in Simvascular are however complex and difficult to understand for physicians. RefCurv could help to translate Simvascular's output into reference curves, which are easy to understand for pediatric cardiologists, for example. We currently work on a connection to the Simvascular framework.\\

Altogether, we can recommend this software for students and researchers of any field, who plan to construct reference curves. Likewise, the software can be used for educational purposes at all levels. For clinicians, this tool can help to understand the underlying methods of the construction of percentile curves and its challenges, such as the tuning of hyperparameters. 


In science and especially in medical science, the usage of proprietary software with restricted access to the code is unfortunately a standard practice. This issue makes it hard for researchers to understand and reproduce the results of other publications. Also, it restricts the scientist from sharing and contributing to other works. This project is entirely open-source and the source code was released under \href{https://choosealicense.com/licenses/gpl-3.0/}{GPLv3}. We encourage other working groups to develop RefCurv and share their knowledge about reference curves so that the scientific community can profit from its value.\\

\newpage
\subsection{Conclusion}
In this paper, we presented RefCurv, a software package enabling to construct reference curves. The software uses the statistical methods of the \pkg{gamlss} package in R and provides a user-friendly GUI for data visualization written in \proglang{Python}. Combining both packages, RefCurv provides a clear structured workflow from data to reference curves. The main features of this software are the model fitting, model selection, sensitivity analysis and model validation.\\

In the present article, we showed exemplarily how RefCurv can improve the application of GAMLSS models. As a result, this package can now also be used by physicians and non-technicians.\\

Due to these advantages, RefCurv could help improving clinical studies to reduce time and costs. We showed how to systematically design studies according to sample size, subject group and medical parameters. In conclusion, a well-designed plan can help to create high-quality reference curves.\\

%% file: appendix.tex
\begin{appendix}

\section{The LMS method by Cole} \label{app:lms}
The LMS method is a special case of a generalized additive model and was originally proposed by \cite{Cole1990}. In summary, the approach can be defined by univariate nonparametric GAM. \\

Let $Y = (y_1, y_2, \dots y_n)^T, \forall y_i > 0$ be a positive random variable with $n$ observations. The explanatory variable is defined by $X = (x_1, x_2, \dots x_n)^T$. The model is defined by the parameters $L$, $M$ and $S$. While $L$ is considered as skewness parameter, $S$ is defined as scale parameter and $M$ location parameter.\\ 
$Y$ should yield a Box-Cox Cole Green (\proglang{BCCG}) distribution denoted by \proglang{BCCG}($M$,$S$,$L$). A transformed random variable $Z$ is given by

\begin{equation} \label{eq:transofrmation}
  Z =\begin{cases}
               \dfrac{1}{L S}\left[\left(\dfrac{Y}{M}\right) ^L -1\right], & \text{if\ } L \ne 0 \\
               \dfrac{1}{S}\ \log \left(\dfrac{Y}{M}\right), & \text{if\ } L = 0 \
            \end{cases}
\end{equation}

for $0<Y<\infty$, where $M>0$, $S>0$ and $-\infty<L<\infty$, and where the random variable $Z$ is assumed to follow a truncated standard normal distribution. \\

\begin{figure}[htbp]
\centering
\includegraphics[width=1\textwidth]{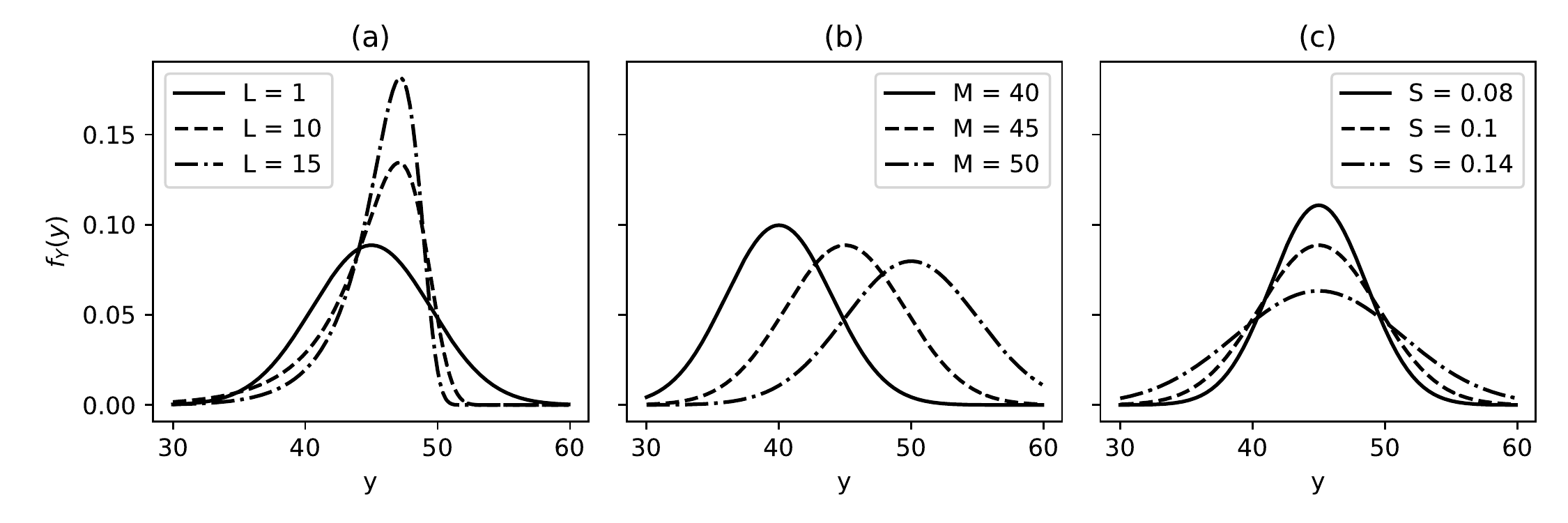}
\caption{\label{fig:BCCG} \textbf{The probability density function $f_Y(y)$ for the BCCG distribution with different values for L, M, and S.} Parameter values: (a) L = 1, M = (40, 45, 50), S = 0.1; (b) L = (1, 10, 15), M = 45, S = 0.1; (c) L = 1, M = 45, S = (0.08, 0.1, 0.14).}
\end{figure}

The probability density function for one observation $y$ and its transform $z$ is given by \\
\begin{equation}\label{eq:pdf}
 f_Y(y) = \dfrac{y^{L-1}\ \exp \left(-\frac{1}{2}z^2\right)}{M^L S \sqrt{2\pi} \Phi\left(\frac{1}{S |L|}\right)}
\end{equation}
where $\Phi()$ is is the cumulative distribution function (cdf) of a standard normal distribution.\\
Figure \ref{fig:BCCG} shows the probability density function for different values of L, M, and S.
\\

Choosing \proglang{BCCG}, the additive model has the form\\
\begin{align}
\begin{split}
M &= h_1(x)\\
\log(S) &= h_2(x)\\
L &= h_3(x)
\end{split}
\end{align}
where $h_i()$ (for $i = 1,2,3$) are non-parametric smoothing functions. Originally, cubic splines \proglang{cs()} have been used as smoothing functions. As alternative to the classic approach, penalized splines were introducted by Eilers and Marx (1996). Penalized Splines (or P-splines) are piecewise polynomials defined by B-spline basis functions in the explanatory variable, where the coefficients of the basis functions are penalized to guarantee sufficient smoothness (Stasinopoulos, 2007). The \proglang(gamlss) package offers the function \proglang{pb()} for fitting penalized splines where \proglang{df} is the desired equivalent number of degrees of freedom.\\ 

The the model with the non-parametric functions $h_k$ ($k = 1,2,3$) is fitted by maximizing the penalized log likelihood function $l_p$, which is defined as
\begin{align}\label{eq:loglikelihood}
\begin{split}
l_p &= l_d - \frac{1}{2}\sum_{k=1}^3 \lambda_k \int_{-\infty}^{\infty}{h''_k(x)} dx\\
	&= l_d  - \frac{1}{2} \lambda_1\int_{-\infty}^{\infty} h''_1(x) dx
			- \frac{1}{2} \lambda_2\int_{-\infty}^{\infty} h''_2(x) dx
			- \frac{1}{2} \lambda_3\int_{-\infty}^{\infty} h''_3(x) dx	
\end{split}	
\end{align}

where $h''_i(x)$ is the second derivative of $h_i(x)$ with respect to $x$. $\lambda_1$, $\lambda_2$, and $\lambda_3$ are smoothing parameters, which have to be predefined.\\
The likelihood function of the data is
\begin{equation}\label{eq:loglikelihood2}
l_d = \sum_{i=1}^n l_i
\end{equation}

and $l_i$ is the log likelihood function of observation $y_i$ which can be computed with (\ref{eq:pdf}).  
The penalized log likelihood function (\ref{eq:loglikelihood}) is maximized iteratively using either the \proglang{RS()} algorithm (\cite{gamlss2005}) or \proglang{CG()} algorithm (Cole and Green), which in turn uses a backfitting algorithm to perform each step of the Fisher scoring procedure.\\
In summary, the LMS method can be applied to a training dataset \proglang{dataset\_training} by using the following piece of code:

\begin{CodeChunk}
\begin{CodeInput}
LMS_model <- gamlss(y ~ pb(x, df = M_df),
		    sigma.formula = ~ pb(x, df = S_df),
		    nu.formula = ~pb(x, df = L_df),
		    family = "BCCG",
		    method = RS(),
		    data = dataset_training)
\end{CodeInput}
\end{CodeChunk}


\newpage
\section{RefCurv - Installation and Software Architecture} \label{app:refcurv}
RefCurv is currently available as version 0.4.2 for Windows (32-bit) and Linux. You can find installation instructions for all systems on \href{https://refcurv.com}{https://refcurv.com}. The source code for each version can be found in the \href{https://github.com/xi2pi/RefCurv}{GitHub respository of RefCurv}.\\ 

For Windows, RefCurv 0.4.2 comes as complete package and does not require any other dependencies to be installed. We tested the software with the versions mentioned below.\\
The main program is written in \proglang{Python} (3.4.0 32-bit) and relies on following packages (with version):
\begin{itemize}
\item numpy (1.14.2)
\item scipy (1.1.0)
\item matplotlib (2.2.2)
\item pandas (0.22.0)
\item PyQt4 (4.11.4)
\end{itemize}
Furthermore, RefCurv is based on \proglang{R} (3.5.2 for 32-bit) and \pkg{gamlss} (5.1-2) add-on package as statistical engine.

\section{Bayesian Information Criterion (BIC)} \label{app:bic}
The Bayesian information criterion (BIC) or Schwarz information criterion (also SIC, SBC, SBIC) is a criterion for model selection. It is typically used to choose among a models with a different setting of hyperparameters. The model with the lowest BIC is preferred.\\
The BIC is defined as

\begin{equation} \label{eq:BIC}
BIC = \ln(n)k - 2 \ln(\hat{l}_d)
\end{equation}

where $\hat{l}_p$ is the maximized value of the likelihood function $l_p$ (\ref{eq:loglikelihood2}). $n$ is the number of observations and $k$ is the number of parameters estimated by the model.\\
The BIC can help to find a compromise between model complexity and goodness of fit. On the one hand, it penalizes high complexity with the term $\ln(n)k$. On the other hand, the goodness of fit is represented as $2 \ln(\hat{l}_d)$. A high goodness of fit will result in a low BIC.  


\section{LMS parameter estimation from percentile curves} \label{app:revcomp}

\begin{figure}[htbp]
\centering
\includegraphics[width=0.8\textwidth]{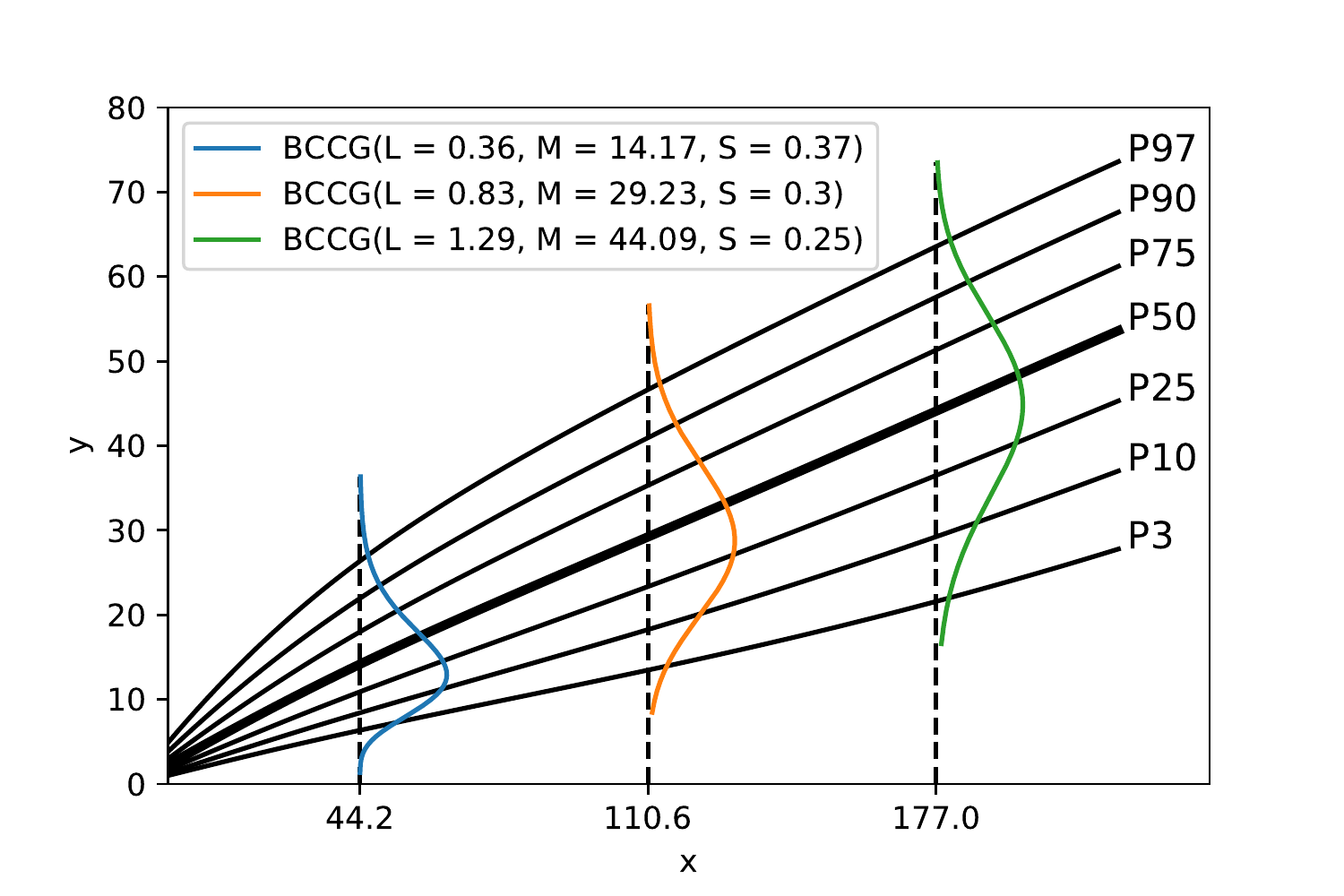}
\caption{\label{fig:lms_estimation} \textbf{LMS parameter estimation from percentile curves.} The BCCG distribution was fitted to the percentile values. The density function for three different positions of the covariate $x = (44.2, 110.6, 177.0)$ are highlighted. }
\end{figure}

\begin{figure}[htbp]
\centering
\includegraphics[width=0.8\textwidth]{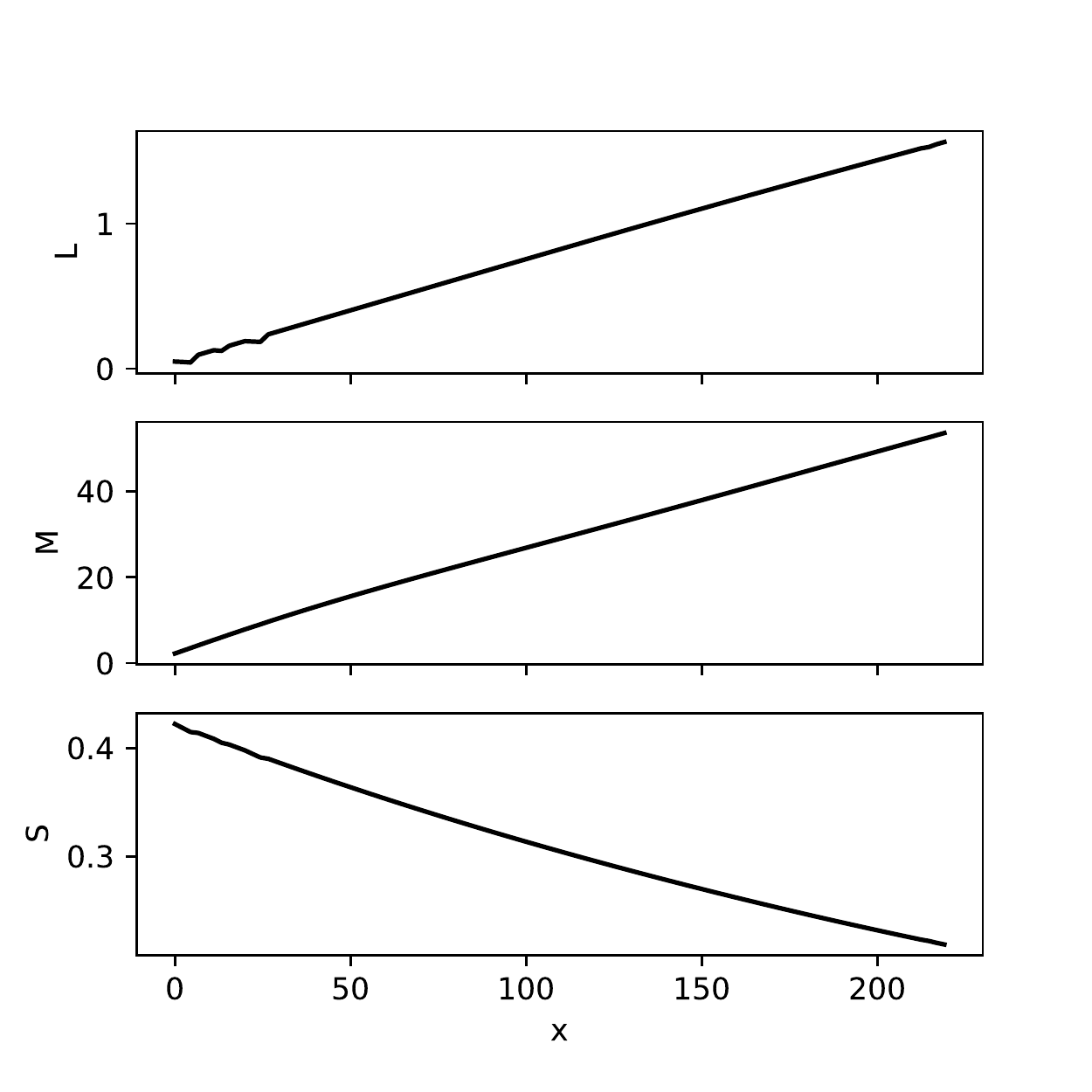}
\caption{\label{fig:lms_estimation_values} \textbf{LMS parameter values against the covariate.} }
\end{figure}

\cite{fenton2007} proposed using Cole's methods to estimate the LMS parameters from percentile curves. They used the Fenton growth chart for preterm infants and generated new percentile curves from the estimated and smoothed LMS parameters. As a result, they found the new curve to be similar to the original curves.\\
This approach can help to use existing charts for z-score prediction of new subjects. Therefore, we implemented an automatized feature to estimate the LMS parameter values for a given chart.\\
Figure \ref{fig:lms_estimation} shows percentile curves and the probability density functions BCCG at three different positions of the covariate $x = (44.2, 110.6, 177.0)$. LMS parameter values were estimated by fitting the probability density function to the percentile curves.
The result of the estimation from percentile curves are L, M, and S over the range of the covariate as shown in figure \ref{fig:lms_estimation_values}. 

\end{appendix}